\documentclass[aps,pre,reprint,superscriptaddress,preprintnumbers,amsmath,amssymb]{revtex4-2}
\usepackage{textcomp}
\usepackage{mathcomp}
\usepackage{amsmath}
\usepackage{graphicx}
\usepackage{dcolumn}
\usepackage{bm}
\usepackage{hyperref}
\usepackage{color}

\begin{document}

\title{Critical scaling for dense granular flow between parallel plates near jamming}

\author{Michio Otsuki}%
 \email{m.otsuki.es@osaka-u.ac.jp}
\affiliation{%
Graduate School of Engineering Science, Osaka University, 1-3 Machikaneyama, Toyonaka, Osaka, 560-8531, Japan
}

\author{Kenta Hayashi}
\affiliation{%
Graduate School of Engineering Science, Osaka University, 1-3 Machikaneyama, Toyonaka, Osaka, 560-8531, Japan
}

\author{Kiwamu Yoshii}
\affiliation{%
Graduate School of Engineering Science, Osaka University, 1-3 Machikaneyama, Toyonaka, Osaka, 560-8531, Japan
}
\affiliation{%
Department of Physics, Nagoya University, Furo-cho, Chikusa-ku, Nagoya, Aichi, 464-8602, Japanķ}

\date{\today}

\begin{abstract}
We numerically study the flow of dense granular materials between parallel plates driven by an external force.
The granular materials form a jammed solid-like state when the external force is below a critical force, while they flow like fluids above the critical force.
The transition is characterized by the mass flux.
The critical force depends on the average packing fraction and the distance between the plates.
The scaling laws for the critical force and the mass flux are predicted theoretically based on a continuum model. 
They are numerically verified.
\end{abstract}

\maketitle

\section{Introduction}


The dynamics of dense amorphous particles such as granular materials, suspensions, foams, and emulsions has been the subject of considerable investigation \cite{Hecke2010, Behringer, Bonn2017}.
These materials can form a solid-like jammed state when the packing fraction $\phi$ exceeds a critical fraction $\phi_{\rm c}$, while they behave like fluids for $\phi<\phi_{\rm c}$.
Such a change in a rheological property is known as the jamming transition. 
The jammed particles above $\phi_{\rm c}$ exhibit an elastic response under small shear stress $\sigma$, but they flow when $\sigma$ exceeds the yield stress $\sigma_{\rm Y}$ \cite{Liu1998}.
In the vicinity of $\phi_{\rm c}$, the pressure, shear modulus, yield stress, and viscosity exhibit continuous transitions, which are characterized by critical power-law behavior \cite{OHern02,OHern2003,Wyart,Zaccone2011,Wyart2005,Hatano2007,Ikeda2012,Kawasaki2015}.
In particular, it has been reported that the flow curve \cite{Olsson2007, Hatano2008, Otsuki2009a,Otsuki2009b,Tighe2010,Nordstrom2010}, complex shear modulus \cite{Tighe11, Otsuki14, Otsuki17}, relaxation modulus \cite{Saitoh2020a}, mean square displacement \cite{Otsuki2012, Saitoh2020b} obey critical scaling laws, similar to those for equilibrium critical phenomena.

Most previous studies of the jamming transition have focused on the critical behavior in homogeneous systems characterized by a constant packing fraction $\phi$ and shear rate $\dot \gamma$.
However, in many manufacturing processes of dense amorphous particles, including granular materials, the formation of a solid-like jammed state is observed in inhomogeneous systems \cite{MiDi2004}. 
This phenomenon manifests in various forms, including clogging \cite{Zuriguel2014,Cai2021}, avalanches \cite{Pouliquen1999, Silbert2003, Forterre2002,Liu2017}, and the separation of fluidized surface layers and frozen bulk regions \cite{Lemieux2000,Komatsu2001,Jop2006,Parker1997,Pignatel2012,Orpe2001,Zheng2019}.
Critical behavior is expected for jamming in these inhomogeneous systems. 
However, little is known about the critical behavior in inhomogeneous systems.

To analyze inhomogeneous granular flows, a continuum theory based on the $\mu (I)$-rheology is useful \cite{MiDi2004}.
In the $\mu (I)$-rheology, a local constitutive equation is assumed, where the bulk friction
\begin{align}
\mu = |\sigma|/p
\label{Def:mu}
\end{align}
with the shear stress $\sigma$ and the normal pressure $p$ is described as a function of the inertia number:
\begin{align}
I = \dot \gamma d /\sqrt{p/\rho_s},
\label{Def:I}
\end{align}
where $\dot \gamma$ is the shear rate, $\rho_{\rm s}$ is the solid grain density, and $d$ is the mean particle diameter \cite{MiDi2004,Cruz2005,Hatano2007b,Peyneau2008,Azema2018,Man2023,Bouzid2013}.
Recent studies have improved the $\mu (I)$-rheology to account for the non-local effect \cite{Kamrin2012,Bouzid2013,Henann2013,Bouzid2015,Zhang2017,Kim2020,Kim2023}.
The $\mu(I)$-rheology is known to reproduce the velocity profiles of various granular flows  \cite{Jop2006,Liu2017,Liu2018,Zheng2019,Barker2022}.
Therefore, we use it to analyze the critical behavior of the inhomogeneous flow in the vicinity of jamming.

In this paper, we numerically study the inhomogeneous steady flow of granular materials driven by an external force between parallel rough plates with the mean packing fraction $\phi_0$.
In Sec. \ref{Sec:setup}, we explain the setup of our model.
Section \ref{Sec:critical} deals with the numerical results of the granular flow.
In Sec. \ref{Sec:theory}, we analytically and numerically investigate scaling laws.
The details of the continuum model are presented in Sec. \ref{Sec:Asetup}.
In Sec. \ref{Sec:static}, the scaling law of the critical force is derived.
In Sec. \ref{Sec:mass}, the critical behavior of the mass flux is studied.
We discuss and conclude our results in Sec. \ref{Sec:discussion}.
In Appendix \ref{App:P}, we present the numerical results on the pressure and estimate the critical fraction.
The details of the derivation of the scaling laws are presented in Appendix \ref{App:analysis}.

\section{Setup}
\label{Sec:setup}

\begin{figure}[h]
    \centering
    \includegraphics[width=0.9\columnwidth]{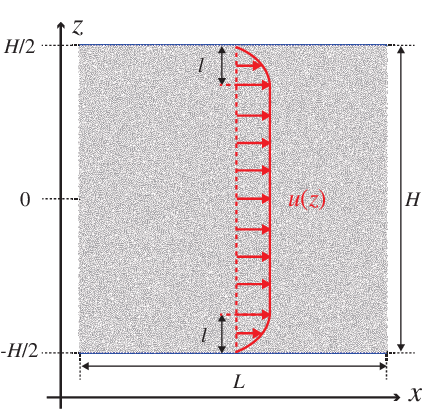}
    \caption{
    Schematic of granular flows between rough parallel plates.
    Red line indicates schematic velocity profiles.
    Blue particles represent the rough plates.
    The size of the fluidized regions are represented by $\ell$.
    }
    \label{setup}
\end{figure}


We consider two-dimensional granular materials consisting of $N$ frictionless particles with identical density $\rho_{\rm s}$ between two parallel rough plates with the mean packing fraction $\phi_0$ as shown in Fig. \ref{setup}.
The plates with length $L$ along the $x$-direction are placed at $z=\pm H/2$ with distance $H$.
We impose periodic boundary conditions in the $x$-direction.
The particles are driven by an external force along the $x$-direction.

The equation of motion is given by
\begin{align}
    m_i \dfrac{d^2}{dt^2} \boldsymbol{r}_i = \sum_{j\neq i} \boldsymbol{F}_{ij} + m_i f \boldsymbol{e}_x,
\end{align}
where $m_i$ and $\boldsymbol{r}_i = (x_i, z_i)$ are the mass and position of particle $i$, respectively.
Here, $f$ is the external force per unit mass, and $\boldsymbol{e}_x=(1,0)$ is the unit vector along the $x$-direction.
The interaction force is given by
\begin{align}
\boldsymbol{F}_{ij} = -\left ( k u^{\rm(n)}_{ij} + \eta v^{\rm(n)}_{ij} \right ) \Theta(d_{ij}-r_{ij}) \boldsymbol{n}_{ij},
\label{Force}
\end{align}
where $\Theta(x)$ is the Heaviside step function satisfying $\Theta(x) = 1$ for $x\ge 0$ and $\Theta(x)=0$ for otherwise.
The elastic constant, viscosity, and diameter of particle $i$ are denoted by k, $\eta$, and $d_i$, respectively.
Here, the relative displacement, the normal relative velocity, and the normal vector are given by
$u^{\rm(n)}_{ij} = r_{ij} - d_{ij}$, $v^{\rm(n)}_{ij} = \dfrac{d}{dt}u^{\rm(n)}_{ij}$, and $\boldsymbol{n}_{ij} = \boldsymbol{r}_{ij}/r_{ij}$, with $d_{ij} = (d_i + d_j)/2$, $\boldsymbol{r}_{ij} = \boldsymbol{r}_i - \boldsymbol{r}_j$, and $r_{ij} = \left |\boldsymbol{r}_{ij} \right |$, respectively.

The granular materials consist of an equal number of particles with diameters $d_0$ and $s d_0$, where $s$ is the size ratio of two types of particles.
The mean diameter is given by $d = (1+s)d_0/2$.
The mean packing fraction of this system is given by $\phi_0 = \pi N(1+s^2)d_0^2/(8LH)$.
The rough plates consist of particles with diameter $d_0$ placed at $(x,z)=(n d_0, \pm H/2)$ with an integer $n$.
We first apply an external force large enough to generate flow, and the external force is changed to a given $f$.
The steady parallel flow is realized after a sufficiently long relaxation time, as shown in Fig. \ref{setup}.
Particles in a long vertical pipe connected to a hopper \cite{Barker2022} and those between pressure-controlled parallel walls \cite{Kamrin2012,Liu2018} exhibit similar parallel flows.


We use $L=200d_0$, $s=1.4$, and $\eta / \sqrt{m_0 k} = 1$, where $m_0$ is the mass of a particle with diameter $d_0$.
The critical packing fraction is estimated as $\phi_{\rm c}\simeq 0.843$ in Appendix \ref{App:P}, below which the particles cannot form the jammed solid-like state.
The parallel flow for $\phi_0 < \phi_{\rm c}$ has been studied in Refs. \cite{Barker2022, Islam2022, Islam2023}, but we focus on dense systems with $\phi_0 \ge \phi_{\rm c}$.
The time evolution equation is solved numerically using LAMMPS (the open source molecular dynamics program from Sandia National Laboratories) with the time step $\Delta t = 0.05 \sqrt{m_0/k}$ \cite{Plimpton1995,thompson2022lammps}.
The units of the velocity, the external force, and the mass flux are $v_0 = d_0 / \sqrt{m_0/k}$, $f_0 = k d_0 / m_0$, and $Q_0 = \sqrt{m_0 k}$, respectively.

\section{Flow near jamming}
\label{Sec:critical}

\begin{figure}[h]
    \centering
    \includegraphics[width=0.9\columnwidth]{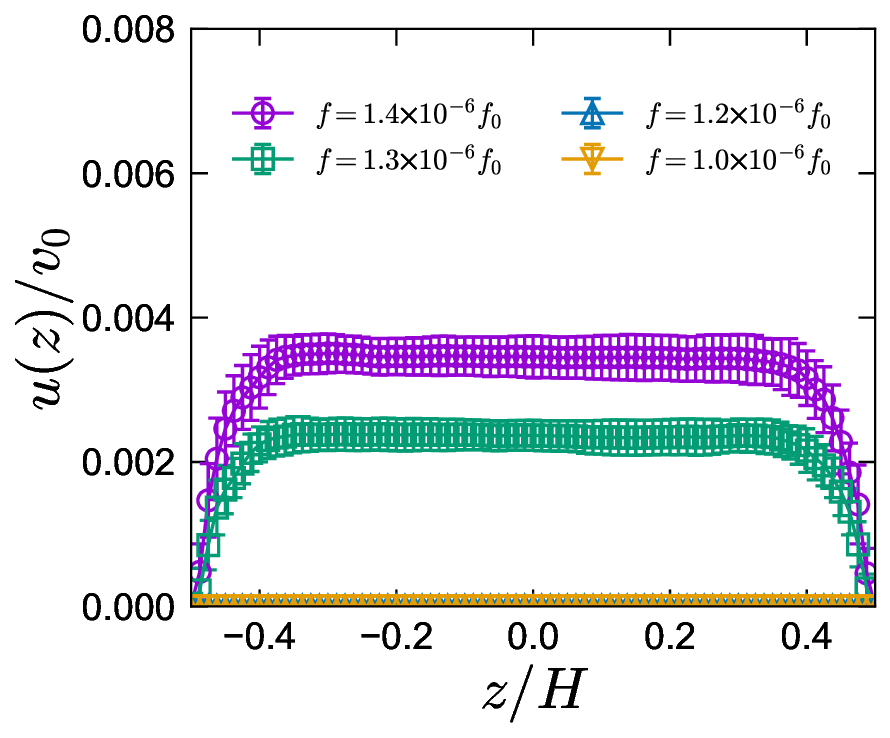}
    \caption{Velocity profile $u(z)$ with $\phi_0=0.850$ and $H=400d_0$ for various values of $f$.}
    \label{Velocity}
\end{figure}

Figure \ref{Velocity} displays the velocity profile $u(z)$ for various values of $f$ with $\phi_0=0.850$ and $H=400d_0$.
Here, $u(z)$ is obtained as the average of the particle velocity $v_{i,x} = \frac{d}{dt} x_i$ at $z$.
The velocity profile is symmetric about $z=0$, and $u(z)=0$ at the rough plates ($z=\pm H/2$).
For $f=1.4\times10^{-6} f_0$ and $f=1.3\times10^{-6} f_0$, a plug flow with a constant velocity is visible near $z=0$.
Near the plates, fluidized regions with nonzero shear rate are observed, where the velocity profile appears parabolic.
For $f=1.2\times10^{-6} f_0$ and $f=1.0\times10^{-6} f_0$, the velocity $u(z)$ is zero in the entire system.
This indicates that the particles are jammed for small $f$.

\begin{figure}
\begin{center}
\includegraphics[width=0.8\linewidth]{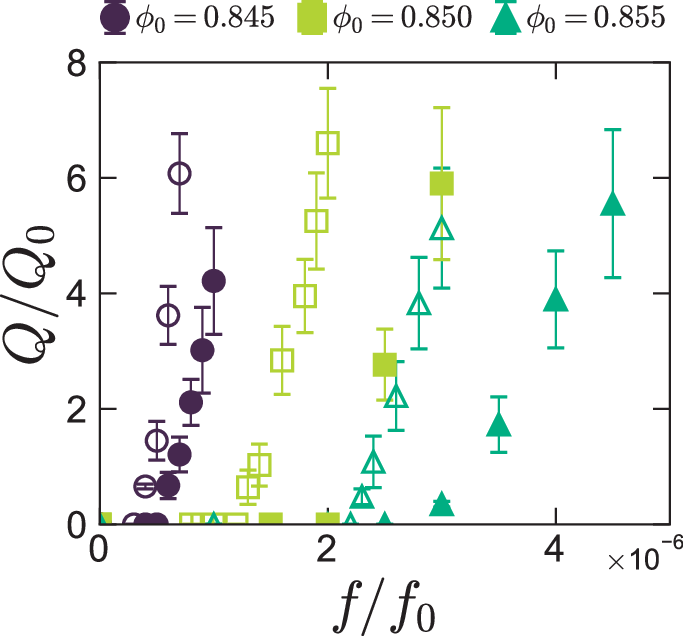}
\end{center}
 \caption{
Mass flux $Q$ against $f$ for various values of $\phi_0$ with $H=300d_0$ (filled symbols) and $H=400d_0$ (open symbols).
 }
 \label{Q_f}
\end{figure}

In Fig. \ref{Q_f}, the mass flux given by $Q = \sum_i m_i v_{i,x}/L$ is plotted against $f$ for various values of $\phi_0$ and $H$.
The mass flux $Q$ increases with $H$ and decreases as $\phi_0$ increases.
For any $\phi_0$ and $H$, $Q$ increases from $0$ as $f$ exceeds a threshold $f_{\rm c}$.
The critical force $f_{\rm c}$ depending on $\phi_0$ and $H$ distinguishes the jammed state with $Q=0$ from the unjammed state with $Q>0$.

In Fig. \ref{phi-phi_cH_VS_f_c}, the critical force $f_{\rm c}$ is plotted against $\phi_0$ for various values of $H$.
We estimate $f_{\rm c}$ as the force $f$ at which $Q$ exceeds $Q_{\rm th}=1.0\times10^{-4}Q_0 (H/d_0)^2$.
Note that $f_{\rm c}$ is unchanged if we choose a smaller $Q_{\rm th}$.
The critical force $f_{\rm c}$ increases as $\phi_0$ exceeds $\phi_{\rm c} \simeq 0.843$ and decreases as $H$ increases.
This indicates that the system jams as $H$ decreases or $\phi_0$ increases.

\begin{figure}[h]
\begin{center}
\includegraphics[width=0.9\linewidth]{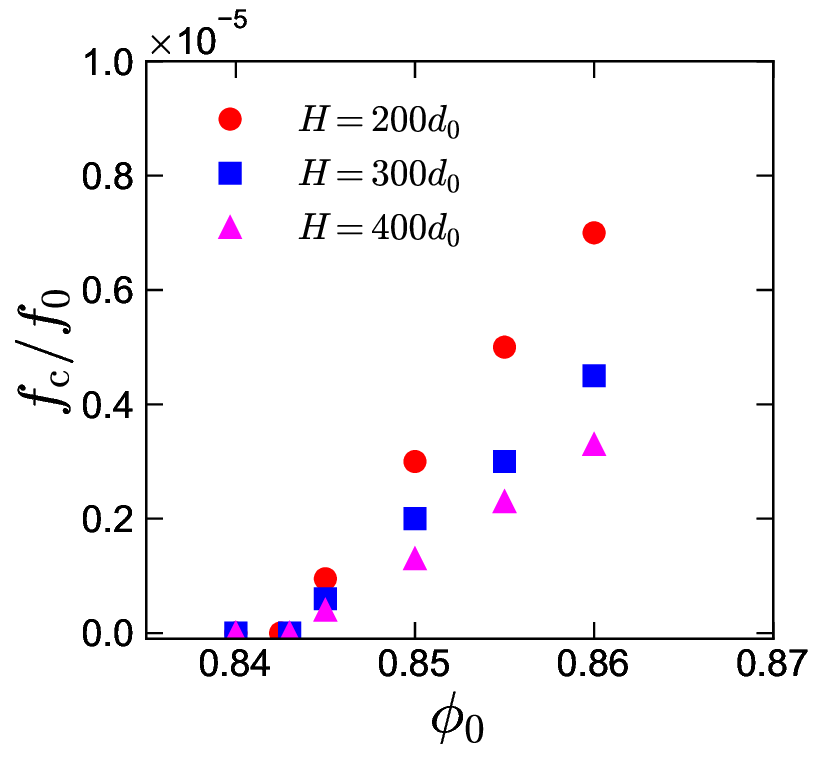}
\end{center}
 \caption{
 Critical force $f_c$ against $\phi_0$ for various values of $H$.
 }
 \label{f_phi}
\end{figure}

\section{Scaling laws near jamming}
\label{Sec:theory}

\subsection{Continuum equations}
\label{Sec:Asetup}

In the steady flow shown in Fig. \ref{Velocity}, the pressure $p(z)$, the shear stress $\sigma(z)$, and the packing fraction $\phi(z)$ at $z$ satisfy the momentum conservation \cite{Evans,Barker2022}
\begin{align}
    & \dfrac{d \sigma(z)}{dz} = - \rho_s \phi(z) f,  \label{Eq:x}\\
    & \dfrac{d p(z)}{dz}=0. \label{Eq:z}
\end{align}
We adopt the $\mu(I)$-rheology \cite{MiDi2004,Cruz2005} as the constitutive equation, where the local inertia number $I(z) = \dot \gamma(z) d /\sqrt{p(z)/\rho_s}$ and bulk friction coefficient $\mu(z)=|\sigma(z)|/p(z)$ satisfy
\begin{align}
I(z) = 
    \begin{cases}
   \mathcal I(\mu(z))  & (\mu(z) > \mu_{\rm s}) \\
    0 & (\mu(z) \le \mu_{\rm s}),
    \end{cases}
\label{Eq:I}
\end{align}
with 
\begin{align}
\mathcal I(\mu) =  I_0 \dfrac{\mu - \mu_{\rm s}}{\mu_2 - \mu},
\label{Eq:mcI}
\end{align}
and the shear rate $\dot \gamma(z) = |d u(z)/dz|$.
Here, $\mu_{\rm s}$ is the static yield coefficient, $\mu_2$ is the maximum value of $\mu$, and $I_0$ is a dimensionless parameter characterizing the non-linear response.
This constitutive equation indicates that the yield stress depends on the pressure, as $\sigma_{\rm Y}=\mu_{\rm s} p$.
We assume that the local pressure $p(z)$ depends on the shear rate $\dot \gamma(z)$ and packing fraction $\phi(z)$: \cite{Hatano2007,Hatano2008,Otsuki2009a,Otsuki2009b,Saitoh16}
\begin{align}
p(z) = \mathcal P(\dot \gamma(z), \phi(z)).
 \label{Eq:p}
\end{align}
The function $\mathcal P(\dot \gamma, \phi)$ satisfies 
\begin{align}
 \lim_{\dot \gamma \to 0} \mathcal P(\dot \gamma, \phi) = \Pi (\phi),
 \label{Eq:p0}
\end{align}
where $\Pi(\phi)$ is the pressure in the quasi-static limit, $\dot \gamma \to 0$.
For particles with the linear repulsive interaction (Eq. \eqref{Force}), $\Pi(\phi)$ obeys a scaling law \cite{OHern02}
\begin{align}
\lim_{\phi \to \phi_{\rm c}} \Pi(\phi) = B |\phi-\phi_{\rm c}|
\label{Lim:Pi}
\end{align}
for $\phi \ge \phi_{\rm c}$ with a constant $B$, which is numerically confirmed in our system as shown in Appendix \ref{App:P}.
Here, we have assumed that $\sigma(z)$ and $p(z)$ are determined by the local quantities $\dot \gamma(z)$ and $\phi(z)$, as the non-local effect is irrelevant for sufficiently large systems \cite{Kamrin2012}.
In fact, we have verified that the same analytical results are obtained for $H/d \gg 1$ even when we apply the non-local constitutive equation used in Ref. \cite{Kamrin2012}.
We have also verified that different forms of $\mathcal I(\mu)$ proposed in Refs. \cite{Hatano2007b,Peyneau2008,Azema2018,Man2023,Bouzid2013} give identical scaling laws for $f_{\rm c}$ and $Q$ shown below.

The mean packing fraction $\phi_0$ satisfies
\begin{align}
\phi_0 = \int_{-H/2}^{H/2} dz \ \phi(z)/H.
\label{Eq:phi}
\end{align}
The boundary condition is given by
\begin{align}
   u(z=\pm H/2) = 0.
   \label{Eq:BC}
\end{align}
The system is symmetric with respect to $z=0$:
\begin{align}
    u(z) = u(-z), \label{Eq:usym} \\
    \sigma(z) = - \sigma(-z). \label{Eq:tausym}
\end{align}
The mass flux $Q$ is given by
\begin{align}
Q = \int_{-H/2}^{H/2} dz \ \rho_s \phi(z) u(z).
\label{Def:Q}
\end{align}

\subsection{Critical force}
\label{Sec:static}

First, we consider the jammed state with $\dot \gamma(z) = 0$ for $f<f_{\rm c}$.
From Eqs. \eqref{Eq:z}, \eqref{Eq:p}, \eqref{Eq:p0}, and \eqref{Eq:phi} with $\dot \gamma(z) = 0$, $\phi(z)$ and $p(z)$ are given by
\begin{align}
    & \phi(z) = \phi_0, \label{phi0}\\
    & p(z) = \Pi (\phi_0), \label{P0}
\end{align}
respectively.
Substituting Eq. \eqref{phi0} into Eq. \eqref{Eq:x} and using Eq. \eqref{Eq:tausym}, we obtain
\begin{align}
    \sigma(z) = -\rho_s \phi_0 f z.
    \label{Eq:tau}
\end{align}
From Eqs. \eqref{P0} and \eqref{Eq:tau}, the bulk friction $\mu(z)=|\sigma (z)|/p(z)$ is given by
\begin{align}
    \mu(z) = \rho_s \phi_0 f |z|/\Pi(\phi_0).
    \label{Eq:mu}
\end{align}

\begin{figure}[h]
\begin{center}
\includegraphics[width=0.9\linewidth]{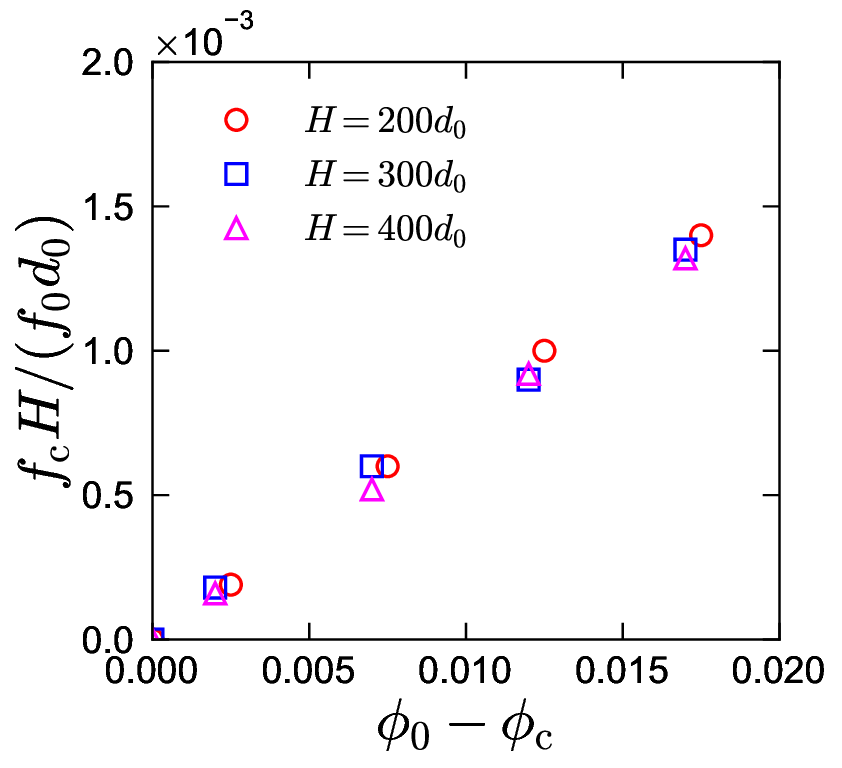}
\end{center}
 \caption{Scaled plot of $f_{\rm c}$ based on Eq. \eqref{Eq:scaling_fc} for various values of $H$.
 }
 \label{phi-phi_cH_VS_f_c}
\end{figure}

According to the constitutive equation, Eq. \eqref{Eq:I}, the jammed state with $\dot \gamma(z) = 0$ is realized when $\mu(z) \le \mu_{\rm s}$ is satisfied for $-H/2 \le z \le H/2$.
Using Eq. \eqref{Eq:mu}, this condition is replaced with $f \le f_{\rm c}$, where
\begin{align}
    f_{\rm c} = \dfrac{2 \mu_{\rm s} \Pi (\phi_0)}{\rho_{\rm s} \phi_0 H}.
    \label{Eq:fc}
\end{align}
For $\phi_0 - \phi_{\rm c} \ll 1$, Eqs. \eqref{Lim:Pi} and \eqref{Eq:fc} give the scaling law
\begin{align}
f_{\rm c} H = \left ( f_0 d_0 \right ) \mathcal F \left (\phi_0 - \phi_{\rm c} \right ),
\label{Eq:scaling_fc}
\end{align}
where $\mathcal F (\xi)$ denotes a scaling function.
In our model, $\mathcal F (\xi)$ is given by $\mathcal F (\xi) = 2 \mu_{\rm s} B \xi / (\rho_{\rm s} f_0 \phi_{\rm c} d_0)$.
In Fig. \ref{phi-phi_cH_VS_f_c}, we present the scaling plot based on Eq. \eqref{Eq:scaling_fc}, which demonstrates the collapse of all data points across various values of $\phi_0$ and $H$. 
This consolidation confirms the validity of the scaling law expressed in Eq. (22).
where the all the data with various $\phi_0$ and $H$ are nicely collapsed.
This confirms the validity of the scaling law, Eq. \eqref{Eq:scaling_fc}.

\subsection{Mass flux}
\label{Sec:mass}


Next, we focus on the critical behavior of the mass flux $Q$ near $f_{\rm c}$.
By introducing a reduced force
\begin{align}
    \epsilon = \dfrac{f-f_c}{f_c},
    \label{Def:ep}
\end{align}
we perturbatively obtain $\phi(z)$, $u(z)$, and $Q$ for $\epsilon \ll 1$ from Eqs. \eqref{Eq:x}--\eqref{Eq:tausym}.
The details of the perturbation theory are shown in Appendix \ref{App:analysis}.
In the lowest order approximation, $\phi(z)$ and $p(z)$ are the same as the solutions for $\dot \gamma = 0$ given by Eqs. \eqref{phi0} and \eqref{P0}.
The local shear rate $\dot \gamma(z)$ is given by
\begin{align}
    \dot \gamma(z) = 
    \begin{cases}
    \dfrac{ \mu_{\rm s} I_0 \sqrt{\Pi(\phi_0)/\rho_{\rm s}}}{ (\mu_{\rm s} - \mu_2) d} \epsilon \zeta(z) & (|z| > H/2 - \ell) \\
    0 & (|z| \le H/2 - \ell),
    \end{cases}
    \label{Eq:dg}
\end{align}
with $\zeta(z) = (|z| - H/2 + \ell)/\ell$. 
The area for $|z| > H/2 - \ell$ represents the fluidized region with size $\ell$ as shown in Fig. \ref{setup}, and $\zeta(z)$ represents the normalized position in the fluidized region.
Size $\ell$ of the fluidized region is given by $\ell = H\epsilon/2$ for $\epsilon \ll 1$.
Integrating the shear rate $\dot \gamma(z)$ with respect to $z$ and using the boundary condition, Eq. \eqref{Eq:BC}, we obtain the velocity profile for $\epsilon \ll 1$ as
\begin{align}
    u(z) = 
    \begin{cases}
    U_{\rm M} \left | 1 -  \{ \zeta(z) \} ^2 \right |  & (|z| > H/2 - \ell) \\
    U_{\rm M} & (|z| \le H/2 - \ell),
    \end{cases}
    \label{Eq:u}
\end{align}
with the maximum velocity
\begin{align}
    U_{\rm M} = \dfrac{ \mu_{\rm s} I_0 }{ 4(\mu_{\rm s} - \mu_2)} 
    \dfrac{ \sqrt{\Pi(\phi_0)/\rho_{\rm s}} H }{ d}  \epsilon^2.
    \label{Eq:umax}
\end{align}
In the fluidized region for $|z| > H/2 - \ell$, the granular materials exhibit a parabolic velocity profile, while a plug flow appears for $|z| \le H/2 - \ell$. 
This velocity profile is similar to that in yield stress fluids \cite{Oldroyd1947,Frigaard1994,Khatib2001}. 
The velocity profile given by Eq. \eqref{Eq:u} is consistent with that in Fig. \ref{Velocity}.
A similar plug flow is obtained for granular materials with $\phi_0 < \phi_c$, but the critical force does not exist in relatively dilute systems \cite{Barker2022, Islam2022, Islam2023}.


The mass flux $Q$ is obtained by substituting the solutions of $\phi(z)$ and $u(z)$ into Eq. \eqref{Def:Q} as 
\begin{align}
Q = \rho_s \phi_0 H U_{\rm M} \label{App:Q}
\end{align}
for $\epsilon \ll 1$.
Substituting Eqs. \eqref{Lim:Pi}, \eqref{Eq:fc}, \eqref{Def:ep}, and \eqref{Eq:umax} into this equation, we obtain a critical scaling law near $\phi_{\rm c}$ for the mass flux $Q$ as
\begin{align}
   \dfrac{Q}{H^2  \sqrt{\phi_0 - \phi_{\rm c}}} = (Q_0/d_0^2) \mathcal Q \left ( \dfrac{f H / (f_0 d_0) }{\phi_0 - \phi_{\rm c}} \right ),
   \label{Eq:scaling}
\end{align}
with a scaling function $\mathcal Q(\xi)$.
In our analytical model, $\mathcal Q(\xi)$ is given by $\mathcal Q(\xi) = \alpha (\beta \xi - 1)^2$, with $\alpha = \dfrac{\mu_{\rm s} I_0 \phi_0 \sqrt{\rho_{\rm s} B}d_0^2}{4 (\mu_{\rm s}-\mu_2)  d Q_0}$ and $\beta = \dfrac{\rho_{\rm s} \phi_0 f_0 d_0}{2 \mu_{\rm s} B}$.

\begin{figure}[h]
\begin{center}
\includegraphics[width=\linewidth]{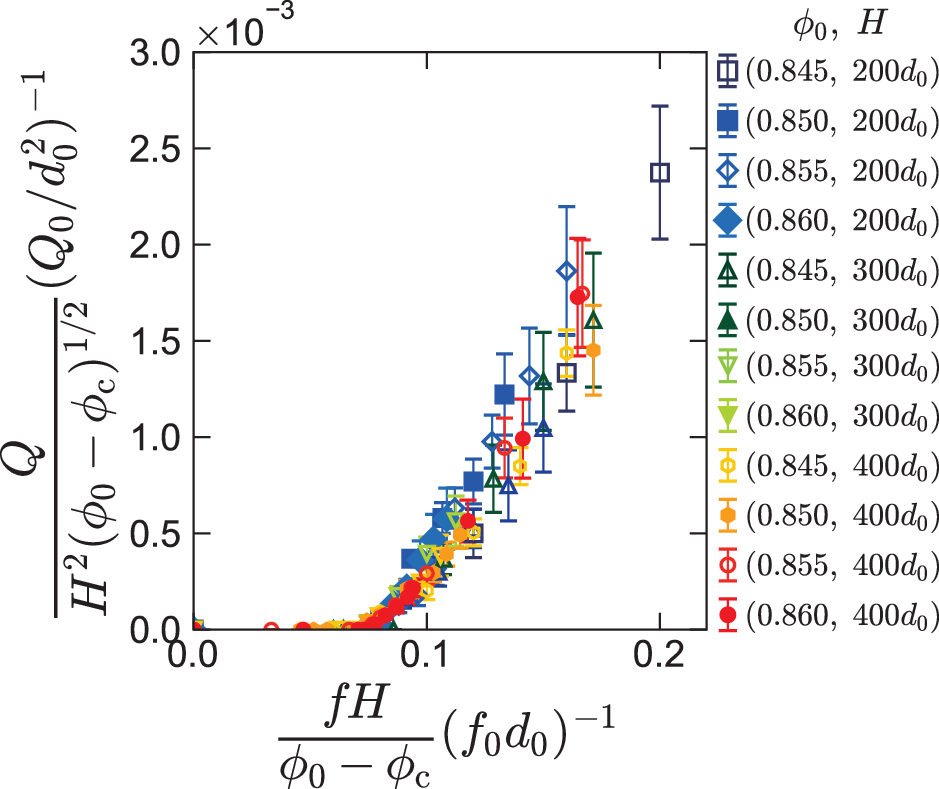}
\end{center}
 \caption{Scaling plot of the mass flux $Q$ based on Eq. \eqref{Eq:scaling} for various values of $\phi_0$ and $H$. The legends represent $(\phi_0, ~H)$. 
 }
 \label{f_c_Q_scalinng}
\end{figure}

In Fig. \ref{f_c_Q_scalinng}, we demonstrate the scaling plot of the mass flux $Q$ based on Eq. \eqref{Eq:scaling}.
Although the system is inhomogeneous, with the shear rate depending on $z$ as shown in Fig. \ref{Velocity}, 
we find excellent consolidation of data in Fig.  \ref{f_c_Q_scalinng}, which mirrors critical scaling laws observed in homogeneous systems near jamming \cite{Olsson2007, Hatano2008, Otsuki2009a,Otsuki2009b,Tighe2010,Nordstrom2010,Tighe11, Otsuki14, Otsuki17,Saitoh2020a,Saitoh2020b}.

\section{Conclusion and discussion}
\label{Sec:discussion}


We have numerically studied the critical behavior of frictionless granular materials between rough parallel plates under an external force.
Jamming is observed as the transition from the jammed state with the mass flux $Q=0$ to the unjammed state with $Q>0$ as $f$ exceeds the critical force $f_{\rm c}$.
Based on a continuum model, we have analytically obtained the velocity profile $u(z)$, which gives the scaling laws for $f_{\rm c}$ and $Q$.
The scaling laws have been numerically verified.
This indicates that critical behaviors
similar to those in homogeneous systems can be
observed in the inhomogeneous system.

In our analysis, we have considered the $z$-dependence of the packing fraction $\phi(z)$, but it is represented as a constant in the lowest order approximation.
In previous studies based on continuum models \cite{Jop2006, Kamrin2012,Liu2017,Liu2018}, the packing fraction is assumed as uniform, but we have confirmed that this assumption is justified for $\epsilon = (f-f_{\rm c})/f_{\rm c} \ll 1$ with $\phi_0 \ge \phi_{\rm c}$.
For relatively dilute systems with $\phi_0 < \phi_{\rm c}$, the critical force $f_{\rm c}$ is zero, and the $z$-dependence of the packing fraction $\phi(z)$ is essential \cite{Barker2022,Islam2022,Islam2023}.

We have considered frictionless particles with a repulsive interaction.
However, we expect that critical behaviors are also observed in particles with friction or attractive interactions, as the $\mu(I)$ rheology is applicable in these systems \cite{Jop2006,Vo2020,Roy2017}. 
An extension of our theory to these systems will be our future work.

\begin{acknowledgments}
The authors thank R. Kuroda, K. Miyazaki, T. Uneyama, N. Oyama, H. Hayakawa, K. Saitoh, S. Takada, and T. Barker for fruitful discussions.
M.O. is partially supported by JSPS KAKENHI (Grant Nos. JP21H01006 and JP23K03248).
K.Y. is partially supported by JSPS Fellows (Grant No.\ 21J13720).
K.Y. would like to thank the Research center for Computational Science, Okazaki, Japan for making its supercomputer system available (Project: 22-IMS-C267 and 23-IMS-C126).
We would like to thank Editage (www.editage.jp) for English language editing.
\end{acknowledgments}

The analytical calculation is performed by K.H and M.O.
The numerical simulations using LAMMPS are performed by K.Y. The first draft of the manuscript was written by M.O.

\appendix

\section{Pressure in the jammed state}
\label{App:P}

\begin{figure}[h]
\begin{center}
\includegraphics[width=0.9\linewidth]{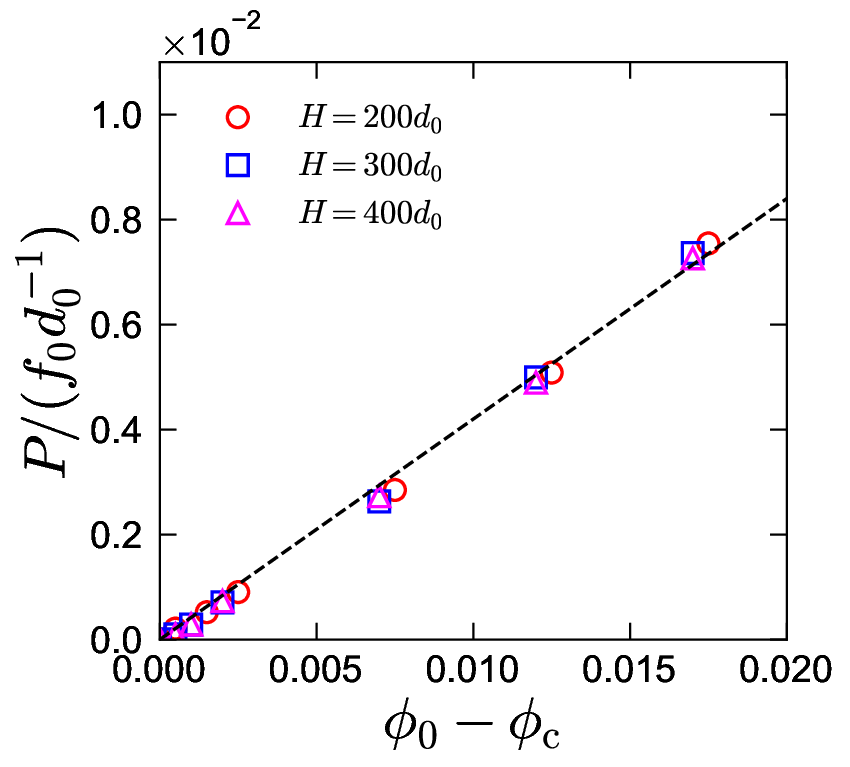}
\end{center}
 \caption{Pressure $P$ at the rough plates with $f=0$ against $\phi_0 - \phi_{\rm c}$ for various values of $H$. 
 The dashed line represents Eq. \eqref{Lim:Pi} with $B=0.4$.
 }
 \label{Pi}
\end{figure}

In this section, we present the numerical results  for the pressure $P$, which is given by $P = \sum_i \sum_{j>i} (x_{ij} F_{ij,x} + y_{ij} F_{ij,y})/(2LH)$ with $\boldsymbol{r}_{ij} = (x_{ij},y_{ij})$ and $\boldsymbol{F}_{ij} = (F_{ij,x},F_{ij,y})$ .
In Fig. \ref{Pi}, we plot the pressure $P$ against $\phi_0 - \phi_{\rm c}$ for $f=0$.
We estimate the critical packing fraction $\phi_{\rm c}$ as $\phi_0$ where $P$ exceeds a threshold $P_{\rm th}=1.0\times10^{-5} f_0 / d_0$ because the nonzero pressure is considered the indicator of jamming \cite{OHern02}.
The critical fraction $\phi_{\rm c}$ has a slight dependence on $H$ as $\phi_{\rm c}= 0.8425, 0.8430,$ and  $0.8430$ for $H/d_0 = 200, 300$, and $400$, respectively.
The pressure $P$ is almost proportional to $\phi_0 - \phi_{\rm c}$ in Fig. \ref{Pi}, which is consistent with Eq. \eqref{Eq:p0}.

\section{Details of analysis}
\label{App:analysis}

In this section, we demonstrate the details of our perturbation theory.
From Eqs. \eqref{Eq:x} and \eqref{Eq:z} with Eq. \eqref{Eq:tausym}, the stress and pressure are given by
\begin{align}
    & \sigma(z) = - \rho_s f \int_0^{z} dz' \ \phi(z'), \label{Eq:taud} \\
    & p(z) = P_{\rm b}, \label{Eq:pd}
\end{align}
where $P_{\rm b}$ is the pressure at the plates. 
Near $z=0$, there is a region where $|\sigma|$ is less than $\mu_{\rm s} P_{\rm b}$.
According to Eq. \eqref{Eq:I}, the shear rate $\dot \gamma$ is zero in this region.
The area for $|z| \le H/2 - \ell$ corresponds to the plug region with $\dot \gamma = 0$, while the area for $|z| > H/2 - \ell$ represents the fluidized region with $\dot \gamma >0$ with the size $\ell$.

With the introduction of the nondimensionalized position 
\begin{align}
    \zeta(z) = (|z| - H/2 + \ell)/\ell,
    \label{Eq:zeta}
\end{align}
we represent the shear rate $\dot \gamma(z)$ as
\begin{align}
    \dot \gamma(z) = 
    \begin{cases}
    \Gamma(\zeta(z)) & (|z| > H/2 - \ell) \\
    0 & (|z| \le H/2 - \ell),
    \end{cases}
    \label{Eq:Ga}
\end{align}
where $\Gamma(\zeta)$ is the shear rate in the fluidized region satisfying 
\begin{align}
\lim_{\zeta \to 0} \Gamma(\zeta) = 0.
\label{C:Ga}
\end{align}
According to Eqs. \eqref{Eq:p}, \eqref{Eq:p0}, \eqref{Eq:pd}, and \eqref{Eq:Ga}, the packing fraction $\phi(z)$ is constant in the plug region.
Hence, $\phi(z)$ is represented as
\begin{align}
    \phi(z) = 
    \begin{cases}
    \phi_{\rm S} \left \{ 1 - \mathcal F (\zeta(z)) \right \}& (|z| > H/2 - \ell) \\
    \phi_{\rm S}  & (|z| \le H/2 - \ell)
    \end{cases}
    \label{Eq:Phi}
\end{align}
with the packing fraction $\phi_{\rm S}$ in the plug region and a function $\mathcal F(\zeta)$
 satisfying 
 \begin{align}
 \lim_{\zeta \to 0} \mathcal F(\zeta) = 0.
 \label{C:F}
 \end{align}

At the boundary of the plug region, $z =  H/2 - \ell$, the shear stress $\sigma$ satisfy $|\sigma|=\mu_s P_{\rm b}$.
Using Eqs. \eqref{Eq:taud}, \eqref{Eq:pd}, and \eqref{Eq:Phi}, this condition is replaced by
\begin{align}
    \mu_s P_{\rm b} = \rho_s \phi_{\rm S} f \left ( \dfrac{H}{2} - \ell\right).
    \label{C:lc}
\end{align}
Substituting \eqref{Eq:Phi} into Eq. \eqref{Eq:phi}, we obtain
\begin{align}
    \phi_0 = \phi_{\rm S} \left \{ 1 - \dfrac{2 \ell}{H} \int_0^1 d \zeta \mathcal F (\zeta) \right \}.
    \label{Eq:PhiS}
\end{align}
From Eqs. \eqref{Eq:taud}, \eqref{Eq:pd}, \eqref{Eq:Phi}, and \eqref{C:lc}, we derive the bulk friction coefficient $\mu = |\sigma|/p$ at $\zeta \ge 0$ as
\begin{align}
    \mu(\zeta) - \mu_s = \mu_s \dfrac{\ell}{H/2 - \ell} \left \{ \zeta - \int_0^{\zeta} d \zeta ' \ \mathcal F ( \zeta ' ) \right \}.
    \label{Eq:muzeta}
\end{align}
Substituting Eqs. \eqref{Eq:pd} and \eqref{Eq:Ga} into Eq. \eqref{Eq:I} with $I = \dot \gamma d / \sqrt{p/\rho_{\rm s}}$, we obtain
\begin{align}
    \dfrac{\Gamma(\zeta) d}{\sqrt{P_{\rm b}/\rho_s}} = \mathcal I \left ( \mu(\zeta) - \mu_s \right ).
    \label{C:G}
\end{align}
Using Eqs. \eqref{Eq:p} and \eqref{Eq:p0} with Eqs. \eqref{Eq:pd}, \eqref{Eq:Ga}, and \eqref{Eq:Phi}, we find
\begin{align}
    & P_{\rm b} = \Pi(\phi_{\rm S}), \label{C:P}\\
    & 0 =  \Delta \left (\phi_{\rm S} \left \{ 1 - \mathcal F (\zeta) \right \},\Gamma(\zeta) \right ), \label{C:FG}
\end{align}
with $\Delta(\phi,\dot \gamma) = P(\phi,\dot \gamma) - \Pi(\phi)$, which satisfies $\lim _{\dot \gamma \to 0} \Delta(\phi,\dot \gamma)= 0$.
From Eqs. \eqref{C:lc}-\eqref{C:FG}, the variables $\ell$, $\phi_{\rm S}$, $P_{\rm b}$, $\Gamma(\zeta)$, and $\mathcal F(\zeta)$ are determined.

For $f \simeq f_{\rm c}$ with $\epsilon \ll 1$, $\ell$, $\phi_{\rm S}$, $P_{\rm b}$, $\Gamma(\zeta)$, and $\mathcal F(\zeta)$ are expanded as
\begin{align}
   \ell & =  \epsilon \ell^{(1)} + O(\epsilon^2), \label{Eq:elle}\\
   \phi_{\rm S} & =  \phi_0 + \epsilon\phi_{\rm S}^{(1)} + O(\epsilon^2), \\
   P_{\rm b} & = \Pi(\phi_0) + \epsilon P_{\rm b}^{(1)} + O(\epsilon^2), \\
   \Gamma(\zeta) & =  \epsilon \Gamma^{(1)}(\zeta)  + O(\epsilon^2), \\
   \mathcal F(\zeta) & =  \epsilon \mathcal F^{(1)}(\zeta)  + O(\epsilon^2). \label{Eq:Fe}
\end{align}
Substituting these equations into Eqs. \eqref{C:lc}-\eqref{C:FG} and extract all terms proportional to $\epsilon$, we obtain 
\begin{align}
    \ell^{(1)} & = H/2, \\
   \phi_{\rm S}^{(1)} & = 0, \\
   P_{\rm b}^{(1)} & =0, \\
   \Gamma^{(1)}(\zeta) & = \dfrac{\mu_{\rm s} I_0 \sqrt{\Pi(\phi_0)}/\rho_{\rm s}}{(\mu_2-\mu_{\rm s})d} \zeta, \\
   \mathcal F^{(1)}(\zeta) & = \dfrac{C_1}{C_2 + C_3}\Gamma^{(1)}(\zeta),
\end{align}
with 
\begin{align}
    C_1 & = \left . \dfrac{\partial \Delta (\phi,\dot \gamma)}{\partial \dot \gamma} \right |_{(\phi,\dot \gamma)=(\phi_0,0)}, \\
    C_2 & = \phi_0 \left . \dfrac{d \Pi (\phi)}{d \phi} \right |_{\phi=\phi_0}, \\
    C_3 & = \left . \phi_0 \dfrac{\partial \Delta (\phi, \dot \gamma)}{\partial \phi}\right |_{(\phi,\dot \gamma)=(\phi_0,0)}.
\end{align}
Substituting these equations into Eqs. \eqref{Eq:elle}-\eqref{Eq:Fe}, we obtain
\begin{align}
   \ell & =  \dfrac{H}{2}\epsilon + O(\epsilon^2), \label{Eq:le} \\
   \phi(z) &= \phi_0 + O(\epsilon), \label{Eq:phie}\\
   p(z) &= \Pi(\phi_0) + O(\epsilon^2), \\
\dot \gamma(z) & = 
    \begin{cases}
    \dfrac{ \mu_{\rm s} I_0 \sqrt{\Pi(\phi_0)/\rho_{\rm s}}}{ (\mu_{\rm s} - \mu_2) d} \epsilon \zeta(z) + O(\epsilon^2) & (|z| > H/2 - \ell) \\
    0 & (|z| \le H/2 - \ell).
    \end{cases}
    \label{Eq:dge}
\end{align}
Neglecting the higher order terms in Eqs. \eqref{Eq:le}-\eqref{Eq:dge}, we obtain $\phi(z)$, $p(z)$, and $\dot \gamma(z)$ as Eqs. \eqref{phi0}, \eqref{P0}, and \eqref{Eq:dg}, respectively.

Integrating $\dot \gamma(z)$ in Eq. \eqref{Eq:dge} with respect to $z$, we obtain
\begin{align}
    u(z) = 
    \begin{cases}
    \tilde U_{\rm M} \epsilon^2 \left | 1 -  \{ \zeta(z) \} ^2 \right |  + O(\epsilon^3) & (|z| > H/2 - \ell) \\
    \tilde U_{\rm M} \epsilon^2  + O(\epsilon^3) & (|z| \le H/2 - \ell)
    \end{cases}
    \label{Eq:ue}
\end{align}
with
\begin{align}
    \tilde U_{\rm M} = \dfrac{ \mu_{\rm s} I_0 }{ 4(\mu_{\rm s} - \mu_2)} 
    \dfrac{ \sqrt{\Pi(\phi_0)/\rho_{\rm s}} H }{ d}.
    \label{Eq:umaxe}
\end{align}
Neglecting $O(\epsilon^3)$ in Eq. \eqref{Eq:ue}, we obtain Eq. \eqref{Eq:u}.
Subsituting Eqs. \eqref{Eq:phie}, \eqref{Eq:ue}, and \eqref{Eq:le} into Eq. \eqref{Def:Q}, we obtain
\begin{align}
    Q = \rho_{\rm s} \phi_0 \tilde U_{\rm M} H \epsilon^2 + O(\epsilon^3).
    \label{Eq:Qe}
\end{align}
Neglecting $O(\epsilon^3)$ in Eq. \eqref{Eq:Qe}, we obtain Eq. \eqref{App:Q}.

\bibliography{export}

\begin{thebibliography}{63}%
\makeatletter
\providecommand \@ifxundefined [1]{%
 \@ifx{#1\undefined}
}%
\providecommand \@ifnum [1]{%
 \ifnum #1\expandafter \@firstoftwo
 \else \expandafter \@secondoftwo
 \fi
}%
\providecommand \@ifx [1]{%
 \ifx #1\expandafter \@firstoftwo
 \else \expandafter \@secondoftwo
 \fi
}%
\providecommand \natexlab [1]{#1}%
\providecommand \enquote  [1]{``#1''}%
\providecommand \bibnamefont  [1]{#1}%
\providecommand \bibfnamefont [1]{#1}%
\providecommand \citenamefont [1]{#1}%
\providecommand \href@noop [0]{\@secondoftwo}%
\providecommand \href [0]{\begingroup \@sanitize@url \@href}%
\providecommand \@href[1]{\@@startlink{#1}\@@href}%
\providecommand \@@href[1]{\endgroup#1\@@endlink}%
\providecommand \@sanitize@url [0]{\catcode `\\12\catcode `\$12\catcode
  `\&12\catcode `\#12\catcode `\^12\catcode `\_12\catcode `\%12\relax}%
\providecommand \@@startlink[1]{}%
\providecommand \@@endlink[0]{}%
\providecommand \url  [0]{\begingroup\@sanitize@url \@url }%
\providecommand \@url [1]{\endgroup\@href {#1}{\urlprefix }}%
\providecommand \urlprefix  [0]{URL }%
\providecommand \Eprint [0]{\href }%
\providecommand \doibase [0]{https://doi.org/}%
\providecommand \selectlanguage [0]{\@gobble}%
\providecommand \bibinfo  [0]{\@secondoftwo}%
\providecommand \bibfield  [0]{\@secondoftwo}%
\providecommand \translation [1]{[#1]}%
\providecommand \BibitemOpen [0]{}%
\providecommand \bibitemStop [0]{}%
\providecommand \bibitemNoStop [0]{.\EOS\space}%
\providecommand \EOS [0]{\spacefactor3000\relax}%
\providecommand \BibitemShut  [1]{\csname bibitem#1\endcsname}%
\let\auto@bib@innerbib\@empty
\bibitem [{\citenamefont {van Hecke}(2010)}]{Hecke2010}%
  \BibitemOpen
  \bibfield  {author} {\bibinfo {author} {\bibfnamefont {M.}~\bibnamefont {van
  Hecke}},\ }\bibfield  {title} {\bibinfo {title} {Jamming of soft particles:
  geometry, mechanics, scaling and isostaticity},\ }\href
  {https://doi.org/10.1088/0953-8984/22/3/033101} {\bibfield  {journal}
  {\bibinfo  {journal} {J. Phys. Condens. Matter}\ }\textbf {\bibinfo {volume}
  {22}},\ \bibinfo {pages} {033101} (\bibinfo {year} {2010})}\BibitemShut
  {NoStop}%
\bibitem [{\citenamefont {Behringer}\ and\ \citenamefont
  {Chakraborty}(2019)}]{Behringer}%
  \BibitemOpen
  \bibfield  {author} {\bibinfo {author} {\bibfnamefont {R.~P.}\ \bibnamefont
  {Behringer}}\ and\ \bibinfo {author} {\bibfnamefont {B.}~\bibnamefont
  {Chakraborty}},\ }\bibfield  {title} {\bibinfo {title} {The physics of
  jamming for granular materials: a review.},\ }\href
  {https://doi.org/10.1088/1361-6633/aadc3c} {\bibfield  {journal} {\bibinfo
  {journal} {Rep. Prog. Phys.}\ }\textbf {\bibinfo {volume} {82}},\ \bibinfo
  {pages} {012601} (\bibinfo {year} {2019})}\BibitemShut {NoStop}%
\bibitem [{\citenamefont {Bonn}\ \emph {et~al.}(2017)\citenamefont {Bonn},
  \citenamefont {Denn}, \citenamefont {Berthier}, \citenamefont {Divoux},\ and\
  \citenamefont {Manneville}}]{Bonn2017}%
  \BibitemOpen
  \bibfield  {author} {\bibinfo {author} {\bibfnamefont {D.}~\bibnamefont
  {Bonn}}, \bibinfo {author} {\bibfnamefont {M.~M.}\ \bibnamefont {Denn}},
  \bibinfo {author} {\bibfnamefont {L.}~\bibnamefont {Berthier}}, \bibinfo
  {author} {\bibfnamefont {T.}~\bibnamefont {Divoux}},\ and\ \bibinfo {author}
  {\bibfnamefont {S.}~\bibnamefont {Manneville}},\ }\bibfield  {title}
  {\bibinfo {title} {Yield stress materials in soft condensed matter},\ }\href
  {https://doi.org/10.1103/RevModPhys.89.035005} {\bibfield  {journal}
  {\bibinfo  {journal} {Rev. Mod. Phys.}\ }\textbf {\bibinfo {volume} {89}},\
  \bibinfo {pages} {035005} (\bibinfo {year} {2017})}\BibitemShut {NoStop}%
\bibitem [{\citenamefont {Liu}\ and\ \citenamefont {Nagel}(1998)}]{Liu1998}%
  \BibitemOpen
  \bibfield  {author} {\bibinfo {author} {\bibfnamefont {A.~J.}\ \bibnamefont
  {Liu}}\ and\ \bibinfo {author} {\bibfnamefont {S.~R.}\ \bibnamefont
  {Nagel}},\ }\bibfield  {title} {\bibinfo {title} {Jamming is not just cool
  any more},\ }\href {https://doi.org/10.1038/23819} {\bibfield  {journal}
  {\bibinfo  {journal} {Nature}\ }\textbf {\bibinfo {volume} {396}},\ \bibinfo
  {pages} {21} (\bibinfo {year} {1998})}\BibitemShut {NoStop}%
\bibitem [{\citenamefont {O'Hern}\ \emph {et~al.}(2002)\citenamefont {O'Hern},
  \citenamefont {Langer}, \citenamefont {Liu},\ and\ \citenamefont
  {Nagel}}]{OHern02}%
  \BibitemOpen
  \bibfield  {author} {\bibinfo {author} {\bibfnamefont {C.~S.}\ \bibnamefont
  {O'Hern}}, \bibinfo {author} {\bibfnamefont {S.~A.}\ \bibnamefont {Langer}},
  \bibinfo {author} {\bibfnamefont {A.~J.}\ \bibnamefont {Liu}},\ and\ \bibinfo
  {author} {\bibfnamefont {S.~R.}\ \bibnamefont {Nagel}},\ }\bibfield  {title}
  {\bibinfo {title} {Random packings of frictionless particles},\ }\href
  {https://doi.org/10.1103/PhysRevLett.88.075507} {\bibfield  {journal}
  {\bibinfo  {journal} {Phys. Rev. Lett.}\ }\textbf {\bibinfo {volume} {88}},\
  \bibinfo {pages} {075507} (\bibinfo {year} {2002})}\BibitemShut {NoStop}%
\bibitem [{\citenamefont {O’Hern}\ \emph {et~al.}(2003)\citenamefont
  {O’Hern}, \citenamefont {Silbert}, \citenamefont {Liu},\ and\ \citenamefont
  {Nagel}}]{OHern2003}%
  \BibitemOpen
  \bibfield  {author} {\bibinfo {author} {\bibfnamefont {C.~S.}\ \bibnamefont
  {O’Hern}}, \bibinfo {author} {\bibfnamefont {L.~E.}\ \bibnamefont
  {Silbert}}, \bibinfo {author} {\bibfnamefont {A.~J.}\ \bibnamefont {Liu}},\
  and\ \bibinfo {author} {\bibfnamefont {S.~R.}\ \bibnamefont {Nagel}},\
  }\bibfield  {title} {\bibinfo {title} {Jamming at zero temperature and zero
  applied stress: The epitome of disorder},\ }\href
  {https://doi.org/10.1103/PhysRevE.68.011306} {\bibfield  {journal} {\bibinfo
  {journal} {Phys. Rev. E}\ }\textbf {\bibinfo {volume} {68}},\ \bibinfo
  {pages} {011306} (\bibinfo {year} {2003})}\BibitemShut {NoStop}%
\bibitem [{\citenamefont {Wyart}(2005)}]{Wyart}%
  \BibitemOpen
  \bibfield  {author} {\bibinfo {author} {\bibfnamefont {M.}~\bibnamefont
  {Wyart}},\ }\bibfield  {title} {\bibinfo {title} {On the rigidity of
  amorphous solids},\ }\href {https://doi.org/10.1051/anphys:2006003}
  {\bibfield  {journal} {\bibinfo  {journal} {Ann. Phys.}\ }\textbf {\bibinfo
  {volume} {30}},\ \bibinfo {pages} {1} (\bibinfo {year} {2005})}\BibitemShut
  {NoStop}%
\bibitem [{\citenamefont {Zaccone}\ and\ \citenamefont
  {Scossa-Romano}(2011)}]{Zaccone2011}%
  \BibitemOpen
  \bibfield  {author} {\bibinfo {author} {\bibfnamefont {A.}~\bibnamefont
  {Zaccone}}\ and\ \bibinfo {author} {\bibfnamefont {E.}~\bibnamefont
  {Scossa-Romano}},\ }\bibfield  {title} {\bibinfo {title} {Approximate
  analytical description of the nonaffine response of amorphous solids},\
  }\href {https://doi.org/10.1103/PhysRevB.83.184205} {\bibfield  {journal}
  {\bibinfo  {journal} {Phys. Rev. B}\ }\textbf {\bibinfo {volume} {83}},\
  \bibinfo {pages} {184205} (\bibinfo {year} {2011})}\BibitemShut {NoStop}%
\bibitem [{\citenamefont {Wyart}\ \emph {et~al.}(2005)\citenamefont {Wyart},
  \citenamefont {Silbert}, \citenamefont {Nagel},\ and\ \citenamefont
  {Witten}}]{Wyart2005}%
  \BibitemOpen
  \bibfield  {author} {\bibinfo {author} {\bibfnamefont {M.}~\bibnamefont
  {Wyart}}, \bibinfo {author} {\bibfnamefont {L.~E.}\ \bibnamefont {Silbert}},
  \bibinfo {author} {\bibfnamefont {S.~R.}\ \bibnamefont {Nagel}},\ and\
  \bibinfo {author} {\bibfnamefont {T.~A.}\ \bibnamefont {Witten}},\ }\bibfield
   {title} {\bibinfo {title} {Effects of compression on the vibrational modes
  of marginally jammed solids},\ }\href
  {https://doi.org/10.1103/PhysRevE.72.051306} {\bibfield  {journal} {\bibinfo
  {journal} {Phys. Rev. E}\ }\textbf {\bibinfo {volume} {72}},\ \bibinfo
  {pages} {051306} (\bibinfo {year} {2005})}\BibitemShut {NoStop}%
\bibitem [{\citenamefont {Hatano}\ \emph {et~al.}(2007)\citenamefont {Hatano},
  \citenamefont {Otsuki},\ and\ \citenamefont {ichi Sasa}}]{Hatano2007}%
  \BibitemOpen
  \bibfield  {author} {\bibinfo {author} {\bibfnamefont {T.}~\bibnamefont
  {Hatano}}, \bibinfo {author} {\bibfnamefont {M.}~\bibnamefont {Otsuki}},\
  and\ \bibinfo {author} {\bibfnamefont {S.}~\bibnamefont {ichi Sasa}},\
  }\bibfield  {title} {\bibinfo {title} {Criticality and scaling relations in a
  sheared granular material},\ }\href {https://doi.org/10.1143/JPSJ.76.023001}
  {\bibfield  {journal} {\bibinfo  {journal} {J. Phys. Soc. Japan}\ }\textbf
  {\bibinfo {volume} {76}},\ \bibinfo {pages} {023001} (\bibinfo {year}
  {2007})}\BibitemShut {NoStop}%
\bibitem [{\citenamefont {Ikeda}\ \emph {et~al.}(2012)\citenamefont {Ikeda},
  \citenamefont {Berthier},\ and\ \citenamefont {Sollich}}]{Ikeda2012}%
  \BibitemOpen
  \bibfield  {author} {\bibinfo {author} {\bibfnamefont {A.}~\bibnamefont
  {Ikeda}}, \bibinfo {author} {\bibfnamefont {L.}~\bibnamefont {Berthier}},\
  and\ \bibinfo {author} {\bibfnamefont {P.}~\bibnamefont {Sollich}},\
  }\bibfield  {title} {\bibinfo {title} {Unified study of glass and jamming
  rheology in soft particle systems},\ }\href
  {https://doi.org/10.1103/PhysRevLett.109.018301} {\bibfield  {journal}
  {\bibinfo  {journal} {Phys. Rev. Lett.}\ }\textbf {\bibinfo {volume} {109}},\
  \bibinfo {pages} {018301} (\bibinfo {year} {2012})}\BibitemShut {NoStop}%
\bibitem [{\citenamefont {Kawasaki}\ \emph {et~al.}(2015)\citenamefont
  {Kawasaki}, \citenamefont {Coslovich}, \citenamefont {Ikeda},\ and\
  \citenamefont {Berthier}}]{Kawasaki2015}%
  \BibitemOpen
  \bibfield  {author} {\bibinfo {author} {\bibfnamefont {T.}~\bibnamefont
  {Kawasaki}}, \bibinfo {author} {\bibfnamefont {D.}~\bibnamefont {Coslovich}},
  \bibinfo {author} {\bibfnamefont {A.}~\bibnamefont {Ikeda}},\ and\ \bibinfo
  {author} {\bibfnamefont {L.}~\bibnamefont {Berthier}},\ }\bibfield  {title}
  {\bibinfo {title} {Diverging viscosity and soft granular rheology in
  non-brownian suspensions},\ }\href
  {https://doi.org/10.1103/PhysRevE.91.012203} {\bibfield  {journal} {\bibinfo
  {journal} {Phys. Rev. E}\ }\textbf {\bibinfo {volume} {91}},\ \bibinfo
  {pages} {012203} (\bibinfo {year} {2015})}\BibitemShut {NoStop}%
\bibitem [{\citenamefont {Olsson}\ and\ \citenamefont
  {Teitel}(2007)}]{Olsson2007}%
  \BibitemOpen
  \bibfield  {author} {\bibinfo {author} {\bibfnamefont {P.}~\bibnamefont
  {Olsson}}\ and\ \bibinfo {author} {\bibfnamefont {S.}~\bibnamefont
  {Teitel}},\ }\bibfield  {title} {\bibinfo {title} {Critical scaling of shear
  viscosity at the jamming transition},\ }\href
  {https://doi.org/10.1103/PhysRevLett.99.178001} {\bibfield  {journal}
  {\bibinfo  {journal} {Phys. Rev. Lett.}\ }\textbf {\bibinfo {volume} {99}},\
  \bibinfo {pages} {178001} (\bibinfo {year} {2007})}\BibitemShut {NoStop}%
\bibitem [{\citenamefont {Hatano}(2008)}]{Hatano2008}%
  \BibitemOpen
  \bibfield  {author} {\bibinfo {author} {\bibfnamefont {T.}~\bibnamefont
  {Hatano}},\ }\bibfield  {title} {\bibinfo {title} {Scaling properties of
  granular rheology near the jamming transition},\ }\href
  {https://doi.org/10.1143/JPSJ.77.123002} {\bibfield  {journal} {\bibinfo
  {journal} {J. Phys. Soc. Japan}\ }\textbf {\bibinfo {volume} {77}},\ \bibinfo
  {pages} {123002} (\bibinfo {year} {2008})}\BibitemShut {NoStop}%
\bibitem [{\citenamefont {Otsuki}\ and\ \citenamefont
  {Hayakawa}(2009{\natexlab{a}})}]{Otsuki2009a}%
  \BibitemOpen
  \bibfield  {author} {\bibinfo {author} {\bibfnamefont {M.}~\bibnamefont
  {Otsuki}}\ and\ \bibinfo {author} {\bibfnamefont {H.}~\bibnamefont
  {Hayakawa}},\ }\bibfield  {title} {\bibinfo {title} {Universal scaling for
  the jamming transition},\ }\href {https://doi.org/10.1143/PTP.121.647}
  {\bibfield  {journal} {\bibinfo  {journal} {Prog. Theor. Phys.}\ }\textbf
  {\bibinfo {volume} {121}},\ \bibinfo {pages} {647} (\bibinfo {year}
  {2009}{\natexlab{a}})}\BibitemShut {NoStop}%
\bibitem [{\citenamefont {Otsuki}\ and\ \citenamefont
  {Hayakawa}(2009{\natexlab{b}})}]{Otsuki2009b}%
  \BibitemOpen
  \bibfield  {author} {\bibinfo {author} {\bibfnamefont {M.}~\bibnamefont
  {Otsuki}}\ and\ \bibinfo {author} {\bibfnamefont {H.}~\bibnamefont
  {Hayakawa}},\ }\bibfield  {title} {\bibinfo {title} {Critical behaviors of
  sheared frictionless granular materials near the jamming transition},\ }\href
  {https://doi.org/10.1103/PhysRevE.80.011308} {\bibfield  {journal} {\bibinfo
  {journal} {Phys. Rev. E}\ }\textbf {\bibinfo {volume} {80}},\ \bibinfo
  {pages} {011308} (\bibinfo {year} {2009}{\natexlab{b}})}\BibitemShut
  {NoStop}%
\bibitem [{\citenamefont {Tighe}\ \emph {et~al.}(2010)\citenamefont {Tighe},
  \citenamefont {Woldhuis}, \citenamefont {Remmers}, \citenamefont {van
  Saarloos},\ and\ \citenamefont {van Hecke}}]{Tighe2010}%
  \BibitemOpen
  \bibfield  {author} {\bibinfo {author} {\bibfnamefont {B.~P.}\ \bibnamefont
  {Tighe}}, \bibinfo {author} {\bibfnamefont {E.}~\bibnamefont {Woldhuis}},
  \bibinfo {author} {\bibfnamefont {J.~J.~C.}\ \bibnamefont {Remmers}},
  \bibinfo {author} {\bibfnamefont {W.}~\bibnamefont {van Saarloos}},\ and\
  \bibinfo {author} {\bibfnamefont {M.}~\bibnamefont {van Hecke}},\ }\bibfield
  {title} {\bibinfo {title} {Model for the scaling of stresses and fluctuations
  in flows near jamming},\ }\href
  {https://doi.org/10.1103/PhysRevLett.105.088303} {\bibfield  {journal}
  {\bibinfo  {journal} {Phys. Rev. Lett.}\ }\textbf {\bibinfo {volume} {105}},\
  \bibinfo {pages} {088303} (\bibinfo {year} {2010})}\BibitemShut {NoStop}%
\bibitem [{\citenamefont {Nordstrom}\ \emph {et~al.}(2010)\citenamefont
  {Nordstrom}, \citenamefont {Verneuil}, \citenamefont {Arratia}, \citenamefont
  {Basu}, \citenamefont {Zhang}, \citenamefont {Yodh}, \citenamefont {Gollub},\
  and\ \citenamefont {Durian}}]{Nordstrom2010}%
  \BibitemOpen
  \bibfield  {author} {\bibinfo {author} {\bibfnamefont {K.~N.}\ \bibnamefont
  {Nordstrom}}, \bibinfo {author} {\bibfnamefont {E.}~\bibnamefont {Verneuil}},
  \bibinfo {author} {\bibfnamefont {P.~E.}\ \bibnamefont {Arratia}}, \bibinfo
  {author} {\bibfnamefont {A.}~\bibnamefont {Basu}}, \bibinfo {author}
  {\bibfnamefont {Z.}~\bibnamefont {Zhang}}, \bibinfo {author} {\bibfnamefont
  {A.~G.}\ \bibnamefont {Yodh}}, \bibinfo {author} {\bibfnamefont {J.~P.}\
  \bibnamefont {Gollub}},\ and\ \bibinfo {author} {\bibfnamefont {D.~J.}\
  \bibnamefont {Durian}},\ }\bibfield  {title} {\bibinfo {title} {Microfluidic
  rheology of soft colloids above and below jamming},\ }\href
  {https://doi.org/10.1103/PhysRevLett.105.175701} {\bibfield  {journal}
  {\bibinfo  {journal} {Phys. Rev. Lett.}\ }\textbf {\bibinfo {volume} {105}},\
  \bibinfo {pages} {175701} (\bibinfo {year} {2010})}\BibitemShut {NoStop}%
\bibitem [{\citenamefont {Tighe}(2011)}]{Tighe11}%
  \BibitemOpen
  \bibfield  {author} {\bibinfo {author} {\bibfnamefont {B.~P.}\ \bibnamefont
  {Tighe}},\ }\bibfield  {title} {\bibinfo {title} {Relaxations and rheology
  near jamming},\ }\href {https://doi.org/10.1103/PhysRevLett.107.158303}
  {\bibfield  {journal} {\bibinfo  {journal} {Phys. Rev. Lett.}\ }\textbf
  {\bibinfo {volume} {107}},\ \bibinfo {pages} {158303} (\bibinfo {year}
  {2011})}\BibitemShut {NoStop}%
\bibitem [{\citenamefont {Otsuki}\ and\ \citenamefont
  {Hayakawa}(2014)}]{Otsuki14}%
  \BibitemOpen
  \bibfield  {author} {\bibinfo {author} {\bibfnamefont {M.}~\bibnamefont
  {Otsuki}}\ and\ \bibinfo {author} {\bibfnamefont {H.}~\bibnamefont
  {Hayakawa}},\ }\bibfield  {title} {\bibinfo {title} {Avalanche contribution
  to shear modulus of granular materials},\ }\href
  {https://doi.org/10.1103/PhysRevE.90.042202} {\bibfield  {journal} {\bibinfo
  {journal} {Phys. Rev. E}\ }\textbf {\bibinfo {volume} {90}},\ \bibinfo
  {pages} {042202} (\bibinfo {year} {2014})}\BibitemShut {NoStop}%
\bibitem [{\citenamefont {Otsuki}\ and\ \citenamefont
  {Hayakawa}(2017)}]{Otsuki17}%
  \BibitemOpen
  \bibfield  {author} {\bibinfo {author} {\bibfnamefont {M.}~\bibnamefont
  {Otsuki}}\ and\ \bibinfo {author} {\bibfnamefont {H.}~\bibnamefont
  {Hayakawa}},\ }\bibfield  {title} {\bibinfo {title} {Discontinuous change of
  shear modulus for frictional jammed granular materials},\ }\href
  {https://doi.org/10.1103/PhysRevE.95.062902} {\bibfield  {journal} {\bibinfo
  {journal} {Phys. Rev. E}\ }\textbf {\bibinfo {volume} {95}},\ \bibinfo
  {pages} {062902} (\bibinfo {year} {2017})}\BibitemShut {NoStop}%
\bibitem [{\citenamefont {Saitoh}\ and\ \citenamefont
  {Kawasaki}(2020)}]{Saitoh2020a}%
  \BibitemOpen
  \bibfield  {author} {\bibinfo {author} {\bibfnamefont {K.}~\bibnamefont
  {Saitoh}}\ and\ \bibinfo {author} {\bibfnamefont {T.}~\bibnamefont
  {Kawasaki}},\ }\bibfield  {title} {\bibinfo {title} {Critical scaling of
  diffusion coefficients and size of rigid clusters of soft athermal particles
  under shear},\ }\href {https://doi.org/10.3389/fphy.2020.00099} {\bibfield
  {journal} {\bibinfo  {journal} {Front. Phys.}\ }\textbf {\bibinfo {volume}
  {8}},\ \bibinfo {pages} {99} (\bibinfo {year} {2020})}\BibitemShut {NoStop}%
\bibitem [{\citenamefont {Otsuki}\ and\ \citenamefont
  {Hayakawa}(2012)}]{Otsuki2012}%
  \BibitemOpen
  \bibfield  {author} {\bibinfo {author} {\bibfnamefont {M.}~\bibnamefont
  {Otsuki}}\ and\ \bibinfo {author} {\bibfnamefont {H.}~\bibnamefont
  {Hayakawa}},\ }\bibfield  {title} {\bibinfo {title} {Critical scaling of a
  jammed system after a quench of temperature},\ }\href
  {https://doi.org/10.1103/PhysRevE.86.031505} {\bibfield  {journal} {\bibinfo
  {journal} {Phys. Rev. E}\ }\textbf {\bibinfo {volume} {86}},\ \bibinfo
  {pages} {031505} (\bibinfo {year} {2012})}\BibitemShut {NoStop}%
\bibitem [{\citenamefont {Saitoh}\ \emph {et~al.}(2020)\citenamefont {Saitoh},
  \citenamefont {Hatano}, \citenamefont {Ikeda},\ and\ \citenamefont
  {Tighe}}]{Saitoh2020b}%
  \BibitemOpen
  \bibfield  {author} {\bibinfo {author} {\bibfnamefont {K.}~\bibnamefont
  {Saitoh}}, \bibinfo {author} {\bibfnamefont {T.}~\bibnamefont {Hatano}},
  \bibinfo {author} {\bibfnamefont {A.}~\bibnamefont {Ikeda}},\ and\ \bibinfo
  {author} {\bibfnamefont {B.~P.}\ \bibnamefont {Tighe}},\ }\bibfield  {title}
  {\bibinfo {title} {Stress relaxation above and below the jamming
  transition},\ }\href {https://doi.org/10.1103/PhysRevLett.124.118001}
  {\bibfield  {journal} {\bibinfo  {journal} {Phys. Rev. Lett.}\ }\textbf
  {\bibinfo {volume} {124}},\ \bibinfo {pages} {118001} (\bibinfo {year}
  {2020})}\BibitemShut {NoStop}%
\bibitem [{\citenamefont {MiDi}(2004)}]{MiDi2004}%
  \BibitemOpen
  \bibfield  {author} {\bibinfo {author} {\bibfnamefont {G.~D.~R.}\
  \bibnamefont {MiDi}},\ }\bibfield  {title} {\bibinfo {title} {On dense
  granular flows},\ }\href {https://doi.org/10.1140/epje/i2003-10153-0}
  {\bibfield  {journal} {\bibinfo  {journal} {Eur. Phys. J. E}\ }\textbf
  {\bibinfo {volume} {14}},\ \bibinfo {pages} {341} (\bibinfo {year}
  {2004})}\BibitemShut {NoStop}%
\bibitem [{\citenamefont {Zuriguel}(2014)}]{Zuriguel2014}%
  \BibitemOpen
  \bibfield  {author} {\bibinfo {author} {\bibfnamefont {I.}~\bibnamefont
  {Zuriguel}},\ }\bibfield  {title} {\bibinfo {title} {Invited review: Clogging
  of granular materials in bottlenecks},\ }\href
  {https://doi.org/10.4279/pip.060014} {\bibfield  {journal} {\bibinfo
  {journal} {Pap. Phys.}\ }\textbf {\bibinfo {volume} {6}},\ \bibinfo {pages}
  {060014} (\bibinfo {year} {2014})}\BibitemShut {NoStop}%
\bibitem [{\citenamefont {Cai}\ \emph {et~al.}(2021)\citenamefont {Cai},
  \citenamefont {Harada},\ and\ \citenamefont {Nordstrom}}]{Cai2021}%
  \BibitemOpen
  \bibfield  {author} {\bibinfo {author} {\bibfnamefont {G.}~\bibnamefont
  {Cai}}, \bibinfo {author} {\bibfnamefont {A.~B.}\ \bibnamefont {Harada}},\
  and\ \bibinfo {author} {\bibfnamefont {K.}~\bibnamefont {Nordstrom}},\
  }\bibfield  {title} {\bibinfo {title} {Mesoscale metrics on approach to the
  clogging point},\ }\href {https://doi.org/10.1007/s10035-021-01133-2}
  {\bibfield  {journal} {\bibinfo  {journal} {Granul. Matter}\ }\textbf
  {\bibinfo {volume} {23}},\ \bibinfo {pages} {69} (\bibinfo {year}
  {2021})}\BibitemShut {NoStop}%
\bibitem [{\citenamefont {Pouliquen}(1999)}]{Pouliquen1999}%
  \BibitemOpen
  \bibfield  {author} {\bibinfo {author} {\bibfnamefont {O.}~\bibnamefont
  {Pouliquen}},\ }\bibfield  {title} {\bibinfo {title} {On the shape of
  granular fronts down rough inclined planes},\ }\href
  {https://doi.org/10.1063/1.870057} {\bibfield  {journal} {\bibinfo  {journal}
  {Phys. Fluids}\ }\textbf {\bibinfo {volume} {11}},\ \bibinfo {pages} {1956}
  (\bibinfo {year} {1999})}\BibitemShut {NoStop}%
\bibitem [{\citenamefont {Silbert}\ \emph {et~al.}(2003)\citenamefont
  {Silbert}, \citenamefont {Landry},\ and\ \citenamefont
  {Grest}}]{Silbert2003}%
  \BibitemOpen
  \bibfield  {author} {\bibinfo {author} {\bibfnamefont {L.~E.}\ \bibnamefont
  {Silbert}}, \bibinfo {author} {\bibfnamefont {J.~W.}\ \bibnamefont
  {Landry}},\ and\ \bibinfo {author} {\bibfnamefont {G.~S.}\ \bibnamefont
  {Grest}},\ }\bibfield  {title} {\bibinfo {title} {Granular flow down a rough
  inclined plane: Transition between thin and thick piles},\ }\href
  {https://doi.org/10.1063/1.1521719} {\bibfield  {journal} {\bibinfo
  {journal} {Phys. Fluids}\ }\textbf {\bibinfo {volume} {15}},\ \bibinfo
  {pages} {1} (\bibinfo {year} {2003})}\BibitemShut {NoStop}%
\bibitem [{\citenamefont {Forterre}\ and\ \citenamefont
  {Pouliquen}(2002)}]{Forterre2002}%
  \BibitemOpen
  \bibfield  {author} {\bibinfo {author} {\bibfnamefont {Y.}~\bibnamefont
  {Forterre}}\ and\ \bibinfo {author} {\bibfnamefont {O.}~\bibnamefont
  {Pouliquen}},\ }\bibfield  {title} {\bibinfo {title} {Stability analysis of
  rapid granular chute flows: formation of longitudinal vortices},\ }\href
  {https://doi.org/10.1017/S0022112002001581} {\bibfield  {journal} {\bibinfo
  {journal} {J. Fluid Mech.}\ }\textbf {\bibinfo {volume} {467}},\ \bibinfo
  {pages} {361} (\bibinfo {year} {2002})}\BibitemShut {NoStop}%
\bibitem [{\citenamefont {Liu}\ and\ \citenamefont {Henann}(2017)}]{Liu2017}%
  \BibitemOpen
  \bibfield  {author} {\bibinfo {author} {\bibfnamefont {D.}~\bibnamefont
  {Liu}}\ and\ \bibinfo {author} {\bibfnamefont {D.~L.}\ \bibnamefont
  {Henann}},\ }\bibfield  {title} {\bibinfo {title} {Non-local continuum
  modelling of steady, dense granular heap flows},\ }\href
  {https://doi.org/10.1017/jfm.2017.554} {\bibfield  {journal} {\bibinfo
  {journal} {J. Fluid Mech.}\ }\textbf {\bibinfo {volume} {831}},\ \bibinfo
  {pages} {212} (\bibinfo {year} {2017})}\BibitemShut {NoStop}%
\bibitem [{\citenamefont {Lemieux}\ and\ \citenamefont
  {Durian}(2000)}]{Lemieux2000}%
  \BibitemOpen
  \bibfield  {author} {\bibinfo {author} {\bibfnamefont {P.-A.}\ \bibnamefont
  {Lemieux}}\ and\ \bibinfo {author} {\bibfnamefont {D.~J.}\ \bibnamefont
  {Durian}},\ }\bibfield  {title} {\bibinfo {title} {From avalanches to fluid
  flow: A continuous picture of grain dynamics down a heap},\ }\href
  {https://doi.org/10.1103/PhysRevLett.85.4273} {\bibfield  {journal} {\bibinfo
   {journal} {Phys. Rev. Lett.}\ }\textbf {\bibinfo {volume} {85}},\ \bibinfo
  {pages} {4273} (\bibinfo {year} {2000})}\BibitemShut {NoStop}%
\bibitem [{\citenamefont {Komatsu}\ \emph {et~al.}(2001)\citenamefont
  {Komatsu}, \citenamefont {Inagaki}, \citenamefont {Nakagawa},\ and\
  \citenamefont {Nasuno}}]{Komatsu2001}%
  \BibitemOpen
  \bibfield  {author} {\bibinfo {author} {\bibfnamefont {T.~S.}\ \bibnamefont
  {Komatsu}}, \bibinfo {author} {\bibfnamefont {S.}~\bibnamefont {Inagaki}},
  \bibinfo {author} {\bibfnamefont {N.}~\bibnamefont {Nakagawa}},\ and\
  \bibinfo {author} {\bibfnamefont {S.}~\bibnamefont {Nasuno}},\ }\bibfield
  {title} {\bibinfo {title} {Creep motion in a granular pile exhibiting steady
  surface flow},\ }\href {https://doi.org/10.1103/PhysRevLett.86.1757}
  {\bibfield  {journal} {\bibinfo  {journal} {Phys. Rev. Lett.}\ }\textbf
  {\bibinfo {volume} {86}},\ \bibinfo {pages} {1757} (\bibinfo {year}
  {2001})}\BibitemShut {NoStop}%
\bibitem [{\citenamefont {Jop}\ \emph {et~al.}(2006)\citenamefont {Jop},
  \citenamefont {Forterre},\ and\ \citenamefont {Pouliquen}}]{Jop2006}%
  \BibitemOpen
  \bibfield  {author} {\bibinfo {author} {\bibfnamefont {P.}~\bibnamefont
  {Jop}}, \bibinfo {author} {\bibfnamefont {Y.}~\bibnamefont {Forterre}},\ and\
  \bibinfo {author} {\bibfnamefont {O.}~\bibnamefont {Pouliquen}},\ }\bibfield
  {title} {\bibinfo {title} {A constitutive law for dense granular flows},\
  }\href {https://doi.org/10.1038/nature04801} {\bibfield  {journal} {\bibinfo
  {journal} {Nature}\ }\textbf {\bibinfo {volume} {441}},\ \bibinfo {pages}
  {727} (\bibinfo {year} {2006})}\BibitemShut {NoStop}%
\bibitem [{\citenamefont {Parker}\ \emph {et~al.}(1997)\citenamefont {Parker},
  \citenamefont {Dijkstra}, \citenamefont {Martin},\ and\ \citenamefont
  {Seville}}]{Parker1997}%
  \BibitemOpen
  \bibfield  {author} {\bibinfo {author} {\bibfnamefont {D.}~\bibnamefont
  {Parker}}, \bibinfo {author} {\bibfnamefont {A.}~\bibnamefont {Dijkstra}},
  \bibinfo {author} {\bibfnamefont {T.}~\bibnamefont {Martin}},\ and\ \bibinfo
  {author} {\bibfnamefont {J.}~\bibnamefont {Seville}},\ }\bibfield  {title}
  {\bibinfo {title} {Positron emission particle tracking studies of spherical
  particle motion in rotating drums},\ }\href
  {https://doi.org/10.1016/S0009-2509(97)00030-4} {\bibfield  {journal}
  {\bibinfo  {journal} {Chem. Eng. Sci.}\ }\textbf {\bibinfo {volume} {52}},\
  \bibinfo {pages} {2011} (\bibinfo {year} {1997})}\BibitemShut {NoStop}%
\bibitem [{\citenamefont {Pignatel}\ \emph {et~al.}(2012)\citenamefont
  {Pignatel}, \citenamefont {Asselin}, \citenamefont {Krieger}, \citenamefont
  {Christov}, \citenamefont {Ottino},\ and\ \citenamefont
  {Lueptow}}]{Pignatel2012}%
  \BibitemOpen
  \bibfield  {author} {\bibinfo {author} {\bibfnamefont {F.}~\bibnamefont
  {Pignatel}}, \bibinfo {author} {\bibfnamefont {C.}~\bibnamefont {Asselin}},
  \bibinfo {author} {\bibfnamefont {L.}~\bibnamefont {Krieger}}, \bibinfo
  {author} {\bibfnamefont {I.~C.}\ \bibnamefont {Christov}}, \bibinfo {author}
  {\bibfnamefont {J.~M.}\ \bibnamefont {Ottino}},\ and\ \bibinfo {author}
  {\bibfnamefont {R.~M.}\ \bibnamefont {Lueptow}},\ }\bibfield  {title}
  {\bibinfo {title} {Parameters and scalings for dry and immersed granular
  flowing layers in rotating tumblers},\ }\href
  {https://doi.org/10.1103/PhysRevE.86.011304} {\bibfield  {journal} {\bibinfo
  {journal} {Phys. Rev. E}\ }\textbf {\bibinfo {volume} {86}},\ \bibinfo
  {pages} {011304} (\bibinfo {year} {2012})}\BibitemShut {NoStop}%
\bibitem [{\citenamefont {Orpe}\ and\ \citenamefont
  {Khakhar}(2001)}]{Orpe2001}%
  \BibitemOpen
  \bibfield  {author} {\bibinfo {author} {\bibfnamefont {A.~V.}\ \bibnamefont
  {Orpe}}\ and\ \bibinfo {author} {\bibfnamefont {D.~V.}\ \bibnamefont
  {Khakhar}},\ }\bibfield  {title} {\bibinfo {title} {Scaling relations for
  granular flow in quasi-two-dimensional rotating cylinders},\ }\href
  {https://doi.org/10.1103/PhysRevE.64.031302} {\bibfield  {journal} {\bibinfo
  {journal} {Phys. Rev. E}\ }\textbf {\bibinfo {volume} {64}},\ \bibinfo
  {pages} {031302} (\bibinfo {year} {2001})}\BibitemShut {NoStop}%
\bibitem [{\citenamefont {Zheng}\ \emph {et~al.}(2019)\citenamefont {Zheng},
  \citenamefont {Bai}, \citenamefont {Yang},\ and\ \citenamefont
  {Yu}}]{Zheng2019}%
  \BibitemOpen
  \bibfield  {author} {\bibinfo {author} {\bibfnamefont {Q.}~\bibnamefont
  {Zheng}}, \bibinfo {author} {\bibfnamefont {L.}~\bibnamefont {Bai}}, \bibinfo
  {author} {\bibfnamefont {L.}~\bibnamefont {Yang}},\ and\ \bibinfo {author}
  {\bibfnamefont {A.}~\bibnamefont {Yu}},\ }\bibfield  {title} {\bibinfo
  {title} {110th anniversary: continuum modeling of granular mixing in a
  rotating drum},\ }\href {https://doi.org/10.1021/acs.iecr.9b03642} {\bibfield
   {journal} {\bibinfo  {journal} {Ind. Eng. Chem.}\ }\textbf {\bibinfo
  {volume} {58}},\ \bibinfo {pages} {19251} (\bibinfo {year}
  {2019})}\BibitemShut {NoStop}%
\bibitem [{\citenamefont {da~Cruz}\ \emph {et~al.}(2005)\citenamefont
  {da~Cruz}, \citenamefont {Emam}, \citenamefont {Prochnow}, \citenamefont
  {Roux},\ and\ \citenamefont {Chevoir}}]{Cruz2005}%
  \BibitemOpen
  \bibfield  {author} {\bibinfo {author} {\bibfnamefont {F.}~\bibnamefont
  {da~Cruz}}, \bibinfo {author} {\bibfnamefont {S.}~\bibnamefont {Emam}},
  \bibinfo {author} {\bibfnamefont {M.}~\bibnamefont {Prochnow}}, \bibinfo
  {author} {\bibfnamefont {J.-N.}\ \bibnamefont {Roux}},\ and\ \bibinfo
  {author} {\bibfnamefont {F.}~\bibnamefont {Chevoir}},\ }\bibfield  {title}
  {\bibinfo {title} {Rheophysics of dense granular materials: Discrete
  simulation of plane shear flows},\ }\href
  {https://doi.org/10.1103/PhysRevE.72.021309} {\bibfield  {journal} {\bibinfo
  {journal} {Phys. Rev. E}\ }\textbf {\bibinfo {volume} {72}},\ \bibinfo
  {pages} {021309} (\bibinfo {year} {2005})}\BibitemShut {NoStop}%
\bibitem [{\citenamefont {Hatano}(2007)}]{Hatano2007b}%
  \BibitemOpen
  \bibfield  {author} {\bibinfo {author} {\bibfnamefont {T.}~\bibnamefont
  {Hatano}},\ }\bibfield  {title} {\bibinfo {title} {Power-law friction in
  closely packed granular materials},\ }\href
  {https://doi.org/10.1103/PhysRevE.75.060301} {\bibfield  {journal} {\bibinfo
  {journal} {Phys. Rev. E}\ }\textbf {\bibinfo {volume} {75}},\ \bibinfo
  {pages} {060301} (\bibinfo {year} {2007})}\BibitemShut {NoStop}%
\bibitem [{\citenamefont {Peyneau}\ and\ \citenamefont
  {Roux}(2008)}]{Peyneau2008}%
  \BibitemOpen
  \bibfield  {author} {\bibinfo {author} {\bibfnamefont {P.-E.}\ \bibnamefont
  {Peyneau}}\ and\ \bibinfo {author} {\bibfnamefont {J.-N.}\ \bibnamefont
  {Roux}},\ }\bibfield  {title} {\bibinfo {title} {Frictionless bead packs have
  macroscopic friction, but no dilatancy},\ }\href
  {https://doi.org/10.1103/PhysRevE.78.011307} {\bibfield  {journal} {\bibinfo
  {journal} {Phys. Rev. E}\ }\textbf {\bibinfo {volume} {78}},\ \bibinfo
  {pages} {011307} (\bibinfo {year} {2008})}\BibitemShut {NoStop}%
\bibitem [{\citenamefont {Émilien Azéma}\ \emph {et~al.}(2018)\citenamefont
  {Émilien Azéma}, \citenamefont {Radjaï},\ and\ \citenamefont
  {Roux}}]{Azema2018}%
  \BibitemOpen
  \bibfield  {author} {\bibinfo {author} {\bibnamefont {Émilien Azéma}},
  \bibinfo {author} {\bibfnamefont {F.}~\bibnamefont {Radjaï}},\ and\ \bibinfo
  {author} {\bibfnamefont {J.-N.}\ \bibnamefont {Roux}},\ }\bibfield  {title}
  {\bibinfo {title} {Inertial shear flow of assemblies of frictionless
  polygons: Rheology and microstructure},\ }\href
  {https://doi.org/10.1140/epje/i2018-11608-9} {\bibfield  {journal} {\bibinfo
  {journal} {Eur. Phys. J. E}\ }\textbf {\bibinfo {volume} {41}},\ \bibinfo
  {pages} {2} (\bibinfo {year} {2018})}\BibitemShut {NoStop}%
\bibitem [{\citenamefont {Man}\ \emph {et~al.}(2023)\citenamefont {Man},
  \citenamefont {Zhang}, \citenamefont {Ge}, \citenamefont {Galindo-Torres},\
  and\ \citenamefont {Hill}}]{Man2023}%
  \BibitemOpen
  \bibfield  {author} {\bibinfo {author} {\bibfnamefont {T.}~\bibnamefont
  {Man}}, \bibinfo {author} {\bibfnamefont {P.}~\bibnamefont {Zhang}}, \bibinfo
  {author} {\bibfnamefont {Z.}~\bibnamefont {Ge}}, \bibinfo {author}
  {\bibfnamefont {S.~A.}\ \bibnamefont {Galindo-Torres}},\ and\ \bibinfo
  {author} {\bibfnamefont {K.~M.}\ \bibnamefont {Hill}},\ }\bibfield  {title}
  {\bibinfo {title} {Friction-dependent rheology of dry granular systems},\
  }\href {https://doi.org/10.1007/s10409-022-22191-x} {\bibfield  {journal}
  {\bibinfo  {journal} {Acta Mech. Sin.}\ }\textbf {\bibinfo {volume} {39}},\
  \bibinfo {pages} {722191} (\bibinfo {year} {2023})}\BibitemShut {NoStop}%
\bibitem [{\citenamefont {Bouzid}\ \emph {et~al.}(2013)\citenamefont {Bouzid},
  \citenamefont {Trulsson}, \citenamefont {Claudin}, \citenamefont {Clément},\
  and\ \citenamefont {Andreotti}}]{Bouzid2013}%
  \BibitemOpen
  \bibfield  {author} {\bibinfo {author} {\bibfnamefont {M.}~\bibnamefont
  {Bouzid}}, \bibinfo {author} {\bibfnamefont {M.}~\bibnamefont {Trulsson}},
  \bibinfo {author} {\bibfnamefont {P.}~\bibnamefont {Claudin}}, \bibinfo
  {author} {\bibfnamefont {E.}~\bibnamefont {Clément}},\ and\ \bibinfo
  {author} {\bibfnamefont {B.}~\bibnamefont {Andreotti}},\ }\bibfield  {title}
  {\bibinfo {title} {Nonlocal rheology of granular flows across yield
  conditions},\ }\href {https://doi.org/10.1103/PhysRevLett.111.238301}
  {\bibfield  {journal} {\bibinfo  {journal} {Phys. Rev. Lett.}\ }\textbf
  {\bibinfo {volume} {111}},\ \bibinfo {pages} {238301} (\bibinfo {year}
  {2013})}\BibitemShut {NoStop}%
\bibitem [{\citenamefont {Kamrin}\ and\ \citenamefont
  {Koval}(2012)}]{Kamrin2012}%
  \BibitemOpen
  \bibfield  {author} {\bibinfo {author} {\bibfnamefont {K.}~\bibnamefont
  {Kamrin}}\ and\ \bibinfo {author} {\bibfnamefont {G.}~\bibnamefont {Koval}},\
  }\bibfield  {title} {\bibinfo {title} {Nonlocal constitutive relation for
  steady granular flow},\ }\href
  {https://doi.org/10.1103/PhysRevLett.108.178301} {\bibfield  {journal}
  {\bibinfo  {journal} {Phys. Rev. Lett.}\ }\textbf {\bibinfo {volume} {108}},\
  \bibinfo {pages} {178301} (\bibinfo {year} {2012})}\BibitemShut {NoStop}%
\bibitem [{\citenamefont {Henann}\ and\ \citenamefont
  {Kamrin}(2013)}]{Henann2013}%
  \BibitemOpen
  \bibfield  {author} {\bibinfo {author} {\bibfnamefont {D.~L.}\ \bibnamefont
  {Henann}}\ and\ \bibinfo {author} {\bibfnamefont {K.}~\bibnamefont
  {Kamrin}},\ }\bibfield  {title} {\bibinfo {title} {A predictive,
  size-dependent continuum model for dense granular flows},\ }\href
  {https://doi.org/10.1073/pnas.1219153110} {\bibfield  {journal} {\bibinfo
  {journal} {Proc. Natl. Acad. Sci. U.S.A}\ }\textbf {\bibinfo {volume}
  {110}},\ \bibinfo {pages} {6730} (\bibinfo {year} {2013})}\BibitemShut
  {NoStop}%
\bibitem [{\citenamefont {Bouzid}\ \emph {et~al.}(2015)\citenamefont {Bouzid},
  \citenamefont {Izzet}, \citenamefont {Trulsson}, \citenamefont {Clément},
  \citenamefont {Claudin},\ and\ \citenamefont {Andreotti}}]{Bouzid2015}%
  \BibitemOpen
  \bibfield  {author} {\bibinfo {author} {\bibfnamefont {M.}~\bibnamefont
  {Bouzid}}, \bibinfo {author} {\bibfnamefont {A.}~\bibnamefont {Izzet}},
  \bibinfo {author} {\bibfnamefont {M.}~\bibnamefont {Trulsson}}, \bibinfo
  {author} {\bibfnamefont {E.}~\bibnamefont {Clément}}, \bibinfo {author}
  {\bibfnamefont {P.}~\bibnamefont {Claudin}},\ and\ \bibinfo {author}
  {\bibfnamefont {B.}~\bibnamefont {Andreotti}},\ }\bibfield  {title} {\bibinfo
  {title} {Non-local rheology in dense granular flows},\ }\href
  {https://doi.org/10.1140/epje/i2015-15125-1} {\bibfield  {journal} {\bibinfo
  {journal} {Eur. Phys. J. E}\ }\textbf {\bibinfo {volume} {38}},\ \bibinfo
  {pages} {125} (\bibinfo {year} {2015})}\BibitemShut {NoStop}%
\bibitem [{\citenamefont {Zhang}\ and\ \citenamefont
  {Kamrin}(2017)}]{Zhang2017}%
  \BibitemOpen
  \bibfield  {author} {\bibinfo {author} {\bibfnamefont {Q.}~\bibnamefont
  {Zhang}}\ and\ \bibinfo {author} {\bibfnamefont {K.}~\bibnamefont {Kamrin}},\
  }\bibfield  {title} {\bibinfo {title} {Microscopic description of the
  granular fluidity field in nonlocal flow modeling},\ }\href
  {https://doi.org/10.1103/PhysRevLett.118.058001} {\bibfield  {journal}
  {\bibinfo  {journal} {Phys. Rev. Lett.}\ }\textbf {\bibinfo {volume} {118}},\
  \bibinfo {pages} {058001} (\bibinfo {year} {2017})}\BibitemShut {NoStop}%
\bibitem [{\citenamefont {Kim}\ and\ \citenamefont {Kamrin}(2020)}]{Kim2020}%
  \BibitemOpen
  \bibfield  {author} {\bibinfo {author} {\bibfnamefont {S.}~\bibnamefont
  {Kim}}\ and\ \bibinfo {author} {\bibfnamefont {K.}~\bibnamefont {Kamrin}},\
  }\bibfield  {title} {\bibinfo {title} {Power-law scaling in granular rheology
  across flow geometries},\ }\href
  {https://doi.org/10.1103/PhysRevLett.125.088002} {\bibfield  {journal}
  {\bibinfo  {journal} {Phys. Rev. Lett.}\ }\textbf {\bibinfo {volume} {125}},\
  \bibinfo {pages} {088002} (\bibinfo {year} {2020})}\BibitemShut {NoStop}%
\bibitem [{\citenamefont {Kim}\ and\ \citenamefont {Kamrin}(2023)}]{Kim2023}%
  \BibitemOpen
  \bibfield  {author} {\bibinfo {author} {\bibfnamefont {S.}~\bibnamefont
  {Kim}}\ and\ \bibinfo {author} {\bibfnamefont {K.}~\bibnamefont {Kamrin}},\
  }\bibfield  {title} {\bibinfo {title} {A second-order non-local model for
  granular flows},\ }\href {https://doi.org/10.3389/fphy.2023.1092233}
  {\bibfield  {journal} {\bibinfo  {journal} {Front. Phys.}\ }\textbf {\bibinfo
  {volume} {11}},\ \bibinfo {pages} {1092233} (\bibinfo {year}
  {2023})}\BibitemShut {NoStop}%
\bibitem [{\citenamefont {Liu}\ and\ \citenamefont {Henann}(2018)}]{Liu2018}%
  \BibitemOpen
  \bibfield  {author} {\bibinfo {author} {\bibfnamefont {D.}~\bibnamefont
  {Liu}}\ and\ \bibinfo {author} {\bibfnamefont {D.~L.}\ \bibnamefont
  {Henann}},\ }\bibfield  {title} {\bibinfo {title} {Size-dependence of the
  flow threshold in dense granular materials},\ }\href
  {https://doi.org/10.1039/C8SM00843D} {\bibfield  {journal} {\bibinfo
  {journal} {Soft Matter}\ }\textbf {\bibinfo {volume} {14}},\ \bibinfo {pages}
  {5294} (\bibinfo {year} {2018})}\BibitemShut {NoStop}%
\bibitem [{\citenamefont {Barker}\ \emph {et~al.}(2022)\citenamefont {Barker},
  \citenamefont {Zhu},\ and\ \citenamefont {Sun}}]{Barker2022}%
  \BibitemOpen
  \bibfield  {author} {\bibinfo {author} {\bibfnamefont {T.}~\bibnamefont
  {Barker}}, \bibinfo {author} {\bibfnamefont {C.}~\bibnamefont {Zhu}},\ and\
  \bibinfo {author} {\bibfnamefont {J.}~\bibnamefont {Sun}},\ }\bibfield
  {title} {\bibinfo {title} {Exact solutions for steady granular flow in
  vertical chutes and pipes},\ }\href {https://doi.org/10.1017/jfm.2021.909}
  {\bibfield  {journal} {\bibinfo  {journal} {J. Fluid Mech.}\ }\textbf
  {\bibinfo {volume} {930}},\ \bibinfo {pages} {A21} (\bibinfo {year}
  {2022})}\BibitemShut {NoStop}%
\bibitem [{\citenamefont {Islam}\ \emph {et~al.}(2022)\citenamefont {Islam},
  \citenamefont {Jenkins},\ and\ \citenamefont {Das}}]{Islam2022}%
  \BibitemOpen
  \bibfield  {author} {\bibinfo {author} {\bibfnamefont {M.~U.}\ \bibnamefont
  {Islam}}, \bibinfo {author} {\bibfnamefont {J.~T.}\ \bibnamefont {Jenkins}},\
  and\ \bibinfo {author} {\bibfnamefont {S.~L.}\ \bibnamefont {Das}},\
  }\bibfield  {title} {\bibinfo {title} {Extended kinetic theory for granular
  flow in a vertical chute},\ }\href {https://doi.org/10.1017/jfm.2022.807}
  {\bibfield  {journal} {\bibinfo  {journal} {J. Fluid Mech.}\ }\textbf
  {\bibinfo {volume} {950}},\ \bibinfo {pages} {A13} (\bibinfo {year}
  {2022})}\BibitemShut {NoStop}%
\bibitem [{\citenamefont {Islam}\ \emph {et~al.}(2023)\citenamefont {Islam},
  \citenamefont {Jenkins},\ and\ \citenamefont {Das}}]{Islam2023}%
  \BibitemOpen
  \bibfield  {author} {\bibinfo {author} {\bibfnamefont {M.~U.}\ \bibnamefont
  {Islam}}, \bibinfo {author} {\bibfnamefont {J.~T.}\ \bibnamefont {Jenkins}},\
  and\ \bibinfo {author} {\bibfnamefont {S.~L.}\ \bibnamefont {Das}},\
  }\bibfield  {title} {\bibinfo {title} {Granular flow through a vertical
  axisymmetric pipe},\ }\href {https://doi.org/10.1103/PhysRevFluids.8.L072301}
  {\bibfield  {journal} {\bibinfo  {journal} {Phys. Rev. Fluids}\ }\textbf
  {\bibinfo {volume} {8}},\ \bibinfo {pages} {L072301} (\bibinfo {year}
  {2023})}\BibitemShut {NoStop}%
\bibitem [{\citenamefont {Plimpton}(1995)}]{Plimpton1995}%
  \BibitemOpen
  \bibfield  {author} {\bibinfo {author} {\bibfnamefont {S.}~\bibnamefont
  {Plimpton}},\ }\bibfield  {title} {\bibinfo {title} {Fast parallel algorithms
  for short-range molecular dynamics},\ }\href
  {https://doi.org/10.1006/jcph.1995.1039} {\bibfield  {journal} {\bibinfo
  {journal} {J. Comput. Phys.}\ }\textbf {\bibinfo {volume} {117}},\ \bibinfo
  {pages} {1} (\bibinfo {year} {1995})}\BibitemShut {NoStop}%
\bibitem [{\citenamefont {Thompson}\ \emph {et~al.}(2022)\citenamefont
  {Thompson}, \citenamefont {Aktulga}, \citenamefont {Berger}, \citenamefont
  {Bolintineanu}, \citenamefont {Brown}, \citenamefont {Crozier}, \citenamefont
  {in't Veld}, \citenamefont {Kohlmeyer}, \citenamefont {Moore}, \citenamefont
  {Nguyen} \emph {et~al.}}]{thompson2022lammps}%
  \BibitemOpen
  \bibfield  {author} {\bibinfo {author} {\bibfnamefont {A.~P.}\ \bibnamefont
  {Thompson}}, \bibinfo {author} {\bibfnamefont {H.~M.}\ \bibnamefont
  {Aktulga}}, \bibinfo {author} {\bibfnamefont {R.}~\bibnamefont {Berger}},
  \bibinfo {author} {\bibfnamefont {D.~S.}\ \bibnamefont {Bolintineanu}},
  \bibinfo {author} {\bibfnamefont {W.~M.}\ \bibnamefont {Brown}}, \bibinfo
  {author} {\bibfnamefont {P.~S.}\ \bibnamefont {Crozier}}, \bibinfo {author}
  {\bibfnamefont {P.~J.}\ \bibnamefont {in't Veld}}, \bibinfo {author}
  {\bibfnamefont {A.}~\bibnamefont {Kohlmeyer}}, \bibinfo {author}
  {\bibfnamefont {S.~G.}\ \bibnamefont {Moore}}, \bibinfo {author}
  {\bibfnamefont {T.~D.}\ \bibnamefont {Nguyen}}, \emph {et~al.},\ }\bibfield
  {title} {\bibinfo {title} {Lammps-a flexible simulation tool for
  particle-based materials modeling at the atomic, meso, and continuum
  scales},\ }\href@noop {} {\bibfield  {journal} {\bibinfo  {journal} {Comput.
  Phys. Commun.}\ }\textbf {\bibinfo {volume} {271}},\ \bibinfo {pages}
  {108171} (\bibinfo {year} {2022})}\BibitemShut {NoStop}%
\bibitem [{\citenamefont {Evans}\ and\ \citenamefont {Morriss}(2008)}]{Evans}%
  \BibitemOpen
  \bibfield  {author} {\bibinfo {author} {\bibfnamefont {D.~J.}\ \bibnamefont
  {Evans}}\ and\ \bibinfo {author} {\bibfnamefont {G.}~\bibnamefont
  {Morriss}},\ }\href {https://doi.org/10.1017/CBO9780511535307} {\emph
  {\bibinfo {title} {Statistical Mechanics of Nonequilibrium Liquids}}}\
  (\bibinfo  {publisher} {Cambridge University Press},\ \bibinfo {year}
  {2008})\BibitemShut {NoStop}%
\bibitem [{\citenamefont {Saitoh}\ and\ \citenamefont
  {Mizuno}(2016)}]{Saitoh16}%
  \BibitemOpen
  \bibfield  {author} {\bibinfo {author} {\bibfnamefont {K.}~\bibnamefont
  {Saitoh}}\ and\ \bibinfo {author} {\bibfnamefont {H.}~\bibnamefont
  {Mizuno}},\ }\bibfield  {title} {\bibinfo {title} {Anomalous energy cascades
  in dense granular materials yielding under simple shear deformations},\
  }\href {https://doi.org/10.1039/C5SM02760H} {\bibfield  {journal} {\bibinfo
  {journal} {Soft Matter}\ }\textbf {\bibinfo {volume} {12}},\ \bibinfo {pages}
  {1360} (\bibinfo {year} {2016})}\BibitemShut {NoStop}%
\bibitem [{\citenamefont {Oldroyd}(1947)}]{Oldroyd1947}%
  \BibitemOpen
  \bibfield  {author} {\bibinfo {author} {\bibfnamefont {J.~G.}\ \bibnamefont
  {Oldroyd}},\ }\bibfield  {title} {\bibinfo {title} {Two-dimensional plastic
  flow of a bingham solid},\ }\href {https://doi.org/10.1017/S0305004100023616}
  {\bibfield  {journal} {\bibinfo  {journal} {Math. Proc. Camb. Philos. Soc.}\
  }\textbf {\bibinfo {volume} {43}},\ \bibinfo {pages} {383} (\bibinfo {year}
  {1947})}\BibitemShut {NoStop}%
\bibitem [{\citenamefont {Frigaard}\ \emph {et~al.}(1994)\citenamefont
  {Frigaard}, \citenamefont {Howison},\ and\ \citenamefont
  {Sobey}}]{Frigaard1994}%
  \BibitemOpen
  \bibfield  {author} {\bibinfo {author} {\bibfnamefont {I.~A.}\ \bibnamefont
  {Frigaard}}, \bibinfo {author} {\bibfnamefont {S.~D.}\ \bibnamefont
  {Howison}},\ and\ \bibinfo {author} {\bibfnamefont {I.~J.}\ \bibnamefont
  {Sobey}},\ }\bibfield  {title} {\bibinfo {title} {On the stability of
  poiseuille flow of a bingham fluid},\ }\href
  {https://doi.org/10.1017/S0022112094004052} {\bibfield  {journal} {\bibinfo
  {journal} {J. Fluid Mech.}\ }\textbf {\bibinfo {volume} {263}},\ \bibinfo
  {pages} {133} (\bibinfo {year} {1994})}\BibitemShut {NoStop}%
\bibitem [{\citenamefont {Khatib}\ and\ \citenamefont
  {Wilson}(2001)}]{Khatib2001}%
  \BibitemOpen
  \bibfield  {author} {\bibinfo {author} {\bibfnamefont {M.~A.}\ \bibnamefont
  {Khatib}}\ and\ \bibinfo {author} {\bibfnamefont {S.}~\bibnamefont
  {Wilson}},\ }\bibfield  {title} {\bibinfo {title} {The development of
  poiseuille flow of a yield-stress fluid},\ }\href
  {https://doi.org/10.1016/S0377-0257(01)00138-0} {\bibfield  {journal}
  {\bibinfo  {journal} {J. Fluid Mech.}\ }\textbf {\bibinfo {volume} {100}},\
  \bibinfo {pages} {1} (\bibinfo {year} {2001})}\BibitemShut {NoStop}%
\bibitem [{\citenamefont {Vo}\ \emph {et~al.}(2020)\citenamefont {Vo},
  \citenamefont {Nezamabadi}, \citenamefont {Mutabaruka}, \citenamefont
  {Delenne},\ and\ \citenamefont {Radjai}}]{Vo2020}%
  \BibitemOpen
  \bibfield  {author} {\bibinfo {author} {\bibfnamefont {T.~T.}\ \bibnamefont
  {Vo}}, \bibinfo {author} {\bibfnamefont {S.}~\bibnamefont {Nezamabadi}},
  \bibinfo {author} {\bibfnamefont {P.}~\bibnamefont {Mutabaruka}}, \bibinfo
  {author} {\bibfnamefont {J.-Y.}\ \bibnamefont {Delenne}},\ and\ \bibinfo
  {author} {\bibfnamefont {F.}~\bibnamefont {Radjai}},\ }\bibfield  {title}
  {\bibinfo {title} {Additive rheology of complex granular flows},\ }\href
  {https://doi.org/10.1038/s41467-020-15263-3} {\bibfield  {journal} {\bibinfo
  {journal} {Nat. Commun.}\ }\textbf {\bibinfo {volume} {11}},\ \bibinfo
  {pages} {1476} (\bibinfo {year} {2020})}\BibitemShut {NoStop}%
\bibitem [{\citenamefont {Roy}\ \emph {et~al.}(2017)\citenamefont {Roy},
  \citenamefont {Luding},\ and\ \citenamefont {Weinhart}}]{Roy2017}%
  \BibitemOpen
  \bibfield  {author} {\bibinfo {author} {\bibfnamefont {S.}~\bibnamefont
  {Roy}}, \bibinfo {author} {\bibfnamefont {S.}~\bibnamefont {Luding}},\ and\
  \bibinfo {author} {\bibfnamefont {T.}~\bibnamefont {Weinhart}},\ }\bibfield
  {title} {\bibinfo {title} {A general(ized) local rheology for wet granular
  materials},\ }\href {https://doi.org/10.1088/1367-2630/aa6141} {\bibfield
  {journal} {\bibinfo  {journal} {New J. Phys.}\ }\textbf {\bibinfo {volume}
  {19}},\ \bibinfo {pages} {043014} (\bibinfo {year} {2017})}\BibitemShut
  {NoStop}%
\end{thebibliography}%
\end{document}